**Original Research**

# Comparative efficacy and safety of pharmacological interventions for the treatment of long COVID in adults: a systematic review and network meta-analysis

**Hailu Zhou[1#], Fei Jiang[1#], and Zhigang Jiang[1*]**

[1]The Laboratary of Heart Development Research, College of Life Science, Hunan Normal University, Changsha 410081, China.

#These authors contributed equally to this work.

*Correspondence should be addressed to

Zhigang Jiang Ph.D.,

The Laboratary of Heart Development Research,

College of Life Science, Hunan Normal University

Changsha 410081, Hunan, China

Email: 201201140149@hunnu.edu.cn

Funded: the Hunan Provincial Natural Science Foundation of China (No. 2023JJ30396)

**Running title:** Systematic Review on Treatment of Long COVID

## Abstract

**Background:** Coronavirus disease 2019 (COVID-19), caused by severe acute respiratory syndrome coronavirus 2 (SARS-CoV-2), is the largest pandemic disease in the past century, and affects humans not only during the acute phase of the infection, but also several weeks to 2 years after the recovery.

**Method:** To estimate the comparative efficacy and safety of pharmacological interventions for the treatment of long COVID in adults, we carried out a systematic review and network meta-analysis (NMA) to evaluate pharmacological interventions and the level of evidence behind each treatment regimen in different clinical settings. Randomized controlled trials (RCTs) and confounding-adjusted observational studies were collected. The efficacy outcomes of interest included all-cause mortality, hospitalization, intensive care unit (ICU) admission, mechanical ventilation (MV) requirement as primary efficacy outcomes, and the recovery of long-COVID symptoms divided into five main categories as secondary efficacy outcomes; the safety outcomes of interest were adverse events (AEs) and serious adverse event (SAEs).

**Results:** We reported odds ratios (ORs) for dichotomous outcomes and standardized mean differences (SMDs) for continuous outcomes with their 95% confidence intervals (CIs), using pairwise and network meta-analysis with random effects. Pairwise meta-analysis results indicated the potential of some traditional COVID-19 drugs in long COVID. The NMA results suggested in post-COVID patients, when compared to control, saline nasal irrigation (SNI) (SMD=21.10, 95%CI [16.91, 25.30]), nitrilotriacetic acid trisodium (NAT) (SMD=7.40, 95%CI [5.79, 9.01]), tetra sodium pyrophosphate (TSPP) (SMD=3.69, 95%CI [2.61, 4.77]) and sodium gluconate (SMD=3.01, 95%CI [1.92, 4.09]) showed significant improvement in anosmia. In terms of thrombosis, rivaroxaban proved to reduce both ATEs (OR=0.33, 95%CI [0.01, 8.19]) and VTEs (OR=0.12, 95%CI [0.01, 0.97]). The therapeutic-dose anticoagulants were associated with a lower risk of ATEs (OR=0.19, 95%CI [0.01, 5.63]) or VTEs (OR=0.04, 95%CI [0.01, 0.38]) but a higher risk of MBEs (OR=1.86, 95%CI [1.19, 2.89]) compared with prophylactic dose or control.

**Conclusions:** The findings of this NMA estimated the comparative efficacy and safety of pharmacological interventions for the treatment of long COVID in adults, providing directions for the development of new drugs against long COVID. Further analysis is required as additional evidence becomes available.

**Keywords**

Long COVID; pharmacological interventions; Systematic review; Network meta-analysis;

## Introduction

Up to one-third of COVID-19 patients [1], even including young adults with mild or asymptomatic disease, might develop post-infection sequelae, and more than 70% of patients hospitalized for COVID-19 might experience at least one persistent symptom [2], often referred to as long COVID [3, 4]. It has been reported that an unwelcome odyssey undertaken by millions of people living with long COVID, which is a complex and sometimes debilitating syndrome that can linger for months or years after an acute SARS-CoV-2 infection [5]. The rapid increase in the number of reports on long-term COVID and the severity of post-infection symptoms are really worrying, making the COVID-19 outbreak a long-term battle with risks beyond our expectations.

## Definition of long COVID

Until now, different descriptions of long COVID have already been put forward. The National Institute for Health and Care Excellence (NICE) have coined the persistent cluster of symptoms as post-COVID syndrome (PCS), which has been further sub-categorized into acute post-COVID syndrome for symptoms persisting 3 weeks beyond initial infection and chronic post-COVID syndrome for symptoms persisting beyond 12 weeks [6]. Besides, the terminology post-acute sequelae of SARS-CoV-2 (PASC) has also been used to define long-term complications of SARS-CoV-2 infection beyond 4 weeks from the onset of symptoms [7]. Listed in the ICD-10 classification since September, 2020, the consensus definition of post-COVID-19 condition (PCC) for adults has been recently proposed by the World Health Organization (WHO) : PCC occurs in individuals with a history of probable or confirmed SARS-CoV-2 infection, usually 3 months from the onset, with symptoms that last for at least 2 months and cannot be explained by an alternative diagnosis [8]. In addition, a wide array of names, including long-haul COVID, long hauler COVID, long-tail COVID, long-term COVID, late sequelae, etc. are equally used to refer to this post-infection sequelae. In this review, the colloquial term long COVID will be used henceforth for consistency.

## Mechanistic understanding of long-COVID symptoms

Since the prevalence and serious health threat of the sequelae in post-COVID-19 patients, it’s very urgent to ramp up the search for long-COVID treatments. However, a key obstacle in the drug design, development and therapy of long COVID is the uncertainty of the underlying cause of the condition. Thus, it's necessary to understand the mechanism of COVID-19 sequelae. Due to the wide distribution pattern of ACE2, the COVID-19 infectious agent, SARS-CoV-2, not only preferentially infects respiratory system causing lasting lung damage [9], but also can widely spread through multiple organs expressing the ACE2 receptor and harbor in several potential tissue repositories, including brain, heart, liver, kidneys, vasculature, gastrointestinal tract, skin and testis [10-13]. In addition, the residual SARS-CoV-2 or its fragments have been reported to be found in body fluids of post-COVID patients, including bronchoalveolar lavage, sputum, saliva, blood, urine, feces, tears, and semen [14, 15]. The proven evidence explains that the multiorgan tropism is a potential factor for the lingering long-term symptoms of COVID-19, where the virus or viral fragments could hide in reservoirs beyond the respiratory tract, which may not be detected by nasopharyngeal swab, and thus continue to cause varying degrees of molecular and cellular damage to multisystem [10, 11].

Several reported hypotheses about the main underlying pathophysiological mechanisms are as follows: 1) delayed viral clearance due to immune exhaustion may trigger chronic COVID-19 symptoms characterized by chronic inflammation and impaired tissue repair [11]; 2) certain SARS-CoV-2-specific antibodies might mistakenly attack host proteins, causing enduring damage long after infection [11, 16]; 3) SARS-CoV-2 could either directly or indirectly induce mitochondrial dysfunction and impaired immunometabolism, leading to the long-term health consequences of COVID-19 [11]; 4) the SARS-CoV-2 infection can alter the composition of gut microbiota, contributing to persistent symptoms of microbiota dysbiosis [11]; 5) imbalance in renin-angiotensin system (RAS) resulting from COVID-19 could cause substantial damage to multiple organ systems, thereby leaving long-term health outcomes in these patients [11]; 6) virus-induced myocardial injury, myocardial inflammation or microvascular thrombosis may be the trigger of persisting cardiac abnormalities, along with limited coronary perfusion or severe hypoxia [17-19]. Therefore, the interactive multi-pathogenesis plays a crucial role in contributing to the complexity of multi-organ sequelae after COVID-19.

**Risk factors for long-COVID symptoms**

The risk factors for developing long COVID are less appreciated. Several studies have done some exploration on patient and clinical characteristics associated with symptoms of long COVID, with risk factors for not returning to usual health as a result of long COVID including older age, female sex, and pre-existing comorbidities including respiratory disease, cardiovascular disease, hypertension, diabetes, obesity and immunosuppressive diseases [20, 21], etc. Besides, a few studies have found an association between the severity of acute COVID-19 infection and post-recovery manifestations in people who have had COVID-19, showing that a more severe acute phase may transform into the development of more severe symptoms of long COVID [22]. It was reported that patients, who suffered from severe COVID-19 in need of prolonged hospitalization, intensive care unit (ICU) admission or mechanical ventilation, were more likely to suffer long-term tissue damage associated with persistent symptoms, with high risk of severe functional disabilities and impaired quality of life [23-25].

In addition, several specific biomarkers are also significant identification risk factors for COVID-19 sequelae, which often reveals potential mechanisms of different types of long COVID symptoms. On one hand, the discovery of inflammatory biomarkers associated with severe COVID-19 provided potential drug targets for the treatment of long-term symptoms. It has been reported that, in severe patients compared to non-severe patients, C-reactive protein, serum amyloid A, interleukin-6, lactate dehydrogenase, neutrophil-to-lymphocyte ratio, D-dimer, cardiac troponin, renal biomarkers showed significantly higher levels, while lymphocytes and platelet count showed significantly lower levels [26]. On the other hand, several inflammatory and vascular biomarkers associated with long COVID remain to be further explored, which also promises to provide potential new approaches for the understanding, diagnosis, and treatment of long-COVID symptoms [27].

**Clinical manifestations of long-COVID symptoms**

As a multisystem disease, long COVID affected different human systems, including: 1) pulmonary system (e.g., cough, dyspnea, respiratory failure, pulmonary thromboembolism, pulmonary embolism, pneumonia, pulmonary vascular damage, pulmonary fibrosis, exercise capacity), 2) nervous system (e.g., loss of taste/smell/hearing, headaches, spasms, convulsions, confusion, visual impairment, nerve pain, dizziness, insomnia, cognitive impairment, brain fog, nausea/vomiting, hemiplegia, ataxia, stroke, cerebral hemorrhage, stress, anxiety, depression, PTSD), 3) cardiovascular system (e.g., vascular hemostasis, blood coagulation, micro/macrovascular thrombosis, thromboembolism, coronary artery atherosclerosis, focal myocardial fibrosis, acute myocardial infarction, myocardial hypertrophy, myocardial inflammation, right ventricular dysfunction, chest pain, palpitations, myocarditis, arrhythmia), 4) liver and kidney system (e.g., acute liver injury, acute kidney injury), 5) gastrointestinal, hepatic and renal systems (e.g., diarrhea, nausea/vomiting, abdominal pain, anorexia, acid reflux, gastrointestinal hemorrhage, lack of appetite/constipation), 6) skin and skeletomuscular system (e.g., hair loss, dermatitis, myalgia, immune-mediated skin diseases, psoriasis, lupus), 7) immune system (e.g., Guillain–Barré syndrome, rheumatoid arthritis, pediatric inflammatory multisystem syndromes such as Kawasaki disease) [11, 18, 19, 28, 29].

## Medication strategy of long-COVID symptoms

Although long-COVID therapies are under way, there is still a long way to go in the search for reliable drug strategies. Until recently, there has been no compelling evidence that any pharmacological treatment directed for long COVID can improve outcomes steadily and safely and has the potential to be put into wide use, meaning that adjuvant antiviral therapy, anti-inflammatory therapy and antithrombotic therapy from the treatment strategies for severe COVID-19 may be the starting point for long-term treatment mechanism for sequelae of COVID-19 [30]. Since the dawn of the post-COVID-19 era, anecdotal evidence, non-peer reviewed articles and strong claims from small clinical trials have put clinicians and patients at risk of using off-label drugs with very low levels of evidence.

A network meta-analysis (NMA) enables a single coherent ranking of numerous interventions, and thus it can aid in decision-makers to choose among a wide range of treatment options [31]. This NMA aimed to provide a comprehensive evaluation of the efficacy and safety of available treatments for patients with long-COVID symptoms including COVID-19 symptoms that may develop into long-term sequelae. We conducted an NMA with selective predefined eligibility criteria for both published and unpublished data and investigated 58 treatment regimens for comparative efficacy and safety. We incorporated 129 studies (109 randomized controlled trials (RCTs) and 20 confounder-adjusted observational studies). The level of certainty behind the evidence for each outcome was also evaluated to assist the decision-making of clinicians and policy makers.

This study comes at a pivotal time, based on global expectations for pharmacological treatments for long COVID and a large number of studies carried out simultaneously over a short period of time. Prospectively designed NMA based on existing and future randomized trials can generate high quality comparative evidence, which can be used to assess drugs potential to be used against long COVID. Therefore, in this study, we aimed to do a systematic review and network meta-analysis of randomized controlled trials (RCTs) and confounding-adjusted observational studies to inform clinical practice and regulatory agencies by comparing different pharmacological interventions versus standard care, placebo or any other intervention for the treatment of COVID-19.

## Materials and Methods

### Search strategy and selection criteria

To decrease subjectivity and the chance of missing the inclusion of studies, we performed an exhaustive online search for eligible studies on PubMed/MEDLINE, Cochrane Library, Embase, CINAHL, PsycINFO and Web of Science for RCTs and observational studies that evaluated treatment responses to pharmacological management in post-COVID patients. As a supplement, we also included several studies of COVID-19 treatment searched through PubMed/MEDLINE, associated with the potential development of pharmacological interventions for sequelae of COVID-19, which we defined into five main categories of symptoms: sensory nerve damage, cardiovascular symptoms, organ failure, psychological disorders and inflammatory response or immune system dysfunction. The complete search strategy is available in Supplementary Appendix 1. Unless translated text could be obtained, non-English studies were excluded. EndNote (version X9, Clarivate, Philadelphia, PA, USA) was used to upload search results and de-duplicate studies by information regarding author, year of publication, title, and reference type. Lastly, we manually searched the lists of references from the identified studies, previous systematic reviews, as well as Google Scholar, to identify additional eligible articles. In addition, ongoing researches on WHO International Clinical Trials Registry Platform, and ClinicalTrials.gov were also hand-searched to acquire latest clinical trials.

## Publication bias

In accordance with the Preferred Reporting Items for Systematic Reviews and Meta-analyses (PRISMA) guidelines [32] for NMA, this systematic review and NMA were conducted from the beginning of 2020 to October 31, 2022. No review protocol or registration details are available.

With enough (≥ 10) studies available for each outcome of interest, publication bias was assessed using the linear regression test of funnel plot asymmetry (Egger's test, implemented using the regtest function in the R metafor package) [33]. A P-value of < 0.05 was considered to suggest the presence of publication bias. When asymmetry was suggested by a visual assessment, we performed exploratory analyses to investigate and adjust for it (trim and fill analysis) using the trimfill function (R metafor package [34]). The magnitude of inconsistency in the study results was assessed by visually examining the forest plot, the L' Abbe plot, the Galbraith radial plot and the graphical display of study heterogeneity (GOSH) plot [35, 36]. Also, subgroup forest maps were plotted to show the impact of different types of treatments and countries on study results. In addition, the Baujat chart was used to analyze the sources of heterogeneity [37]. To diagnose whether the model is reasonable, we used the quantile-quantile plot (QQ plot) [38] to check the normality of the residual. In our NMA, the net heat plot was constructed using R corrplot package to visualize the inconsistency matrix and detect specific comparisons which introduced large inconsistencies.

## Evaluation of outcomes

The efficacy outcomes of interest included all-cause mortality, hospital admission, hospital days, intensive care unit (ICU) admission, intensive care unit (ICU) days, mechanical ventilation (MV) requirement and mechanical ventilation (MV) days as primary efficacy outcomes, and the recovery of COVID-19 symptoms characterized by improvement of smell scores, venous thrombotic events (VTEs) and arterial thrombotic events (ATEs) as secondary efficacy outcomes; the safety outcomes of interest were adverse events (AEs) and serious adverse event (SAEs) as primary safety outcomes, and major bleeding events (MBEs).

## Data extraction

The basic information of references screened for full text were exported from EndNote. Two authors independently full-text filtered the retrieved references and extracted the data, using a predefined data extraction sheet. For each eligible study, we collected trial characteristics, interventions, demographic characteristics and outcomes of interest. For binary outcomes of interest, numbers of events and total numbers of patients were collected. For continuous data, mean, standard deviation of outcomes and total numbers of patients were collected. Two reviewers resolved discrepancies via discussion and a third party adjudicated if any conflict arose. For multiple reports on the same trial, we adopted the latest peer-reviewed publication. Participants, interventions, comparisons, outcomes, and study design (PICOS) of all included studies are described in Supplementary Appendix 2.

## Risk of bias assessment

Risk of bias for each included randomized controlled trial (RCT) was assessed using the Collaboration Risk of Bias Tool [39, 40], where potential sources of bias include random sequence generation, allocation concealment, blinding of participants and staff, blinding of outcome assessors, incomplete outcome data, and selective reporting. Each trial received a study level score of low, high, or unclear risk of bias for each domain. Two authors independently conducted this assessment, and discrepancies were resolved by consensus. Again, one reviewer evaluated all records with a second reviewer independently evaluating 20% of the records between them, and disagreements being resolved by consensus.

## Data synthesis

In contrast to pairwise meta-analysis, which allows direct comparison of one intervention to another based on head-to-head data from randomized trials, network meta-analysis (NMA) extends the principles of meta-analysis to evaluate multiple treatments in a single analysis by combining direct and indirect evidence, facilitating simultaneous comparison of the efficacy or safety of multiple interventions [41, 42]. Included In our NMA, several multi-arm RCTs reported results for different interventions versus control (placebo/SOC), were divided into separated two-arm studies in pairwise meta-analysis. In the primary analysis, the control group with the largest number of participants was used if the periods of patient enrollment of specified interventions had overlaps, otherwise we considered a new control group which combined the control groups of all studies for the same RCT. As observational studies are more vulnerable to bias, we included only the studies that accounted for relevant confounding variables by showing that baseline characteristics were similar ($p > 0.05$ for baseline characteristics) or through methods such as propensity score matching (PSM) [43], subgroup analyses, or regression model adjustment. Effect estimates were pooled by way of random-effects meta-analyses (inverse-variance method for effect size, restricted maximum-likelihood estimator for variance) [44] using R version 4.2.3. We reported odds ratios (ORs) [45] for dichotomous outcomes and standardized mean differences (SMDs) [46] for continuous outcomes with their 95% confidence intervals (CIs), using pairwise and network meta-analysis with random effects. Both unadjusted and adjusted estimates were extracted and pooled separately. When median values and ranges/interquartile ranges were provided, they were used to estimate the mean values and standard deviations [47]. If necessary, means and standard deviations were combined using formulae available in the Cochrane Handbook [39]. The rank of effect or safety estimation for each treatment was investigated using the surface under the cumulative rank curve (SUCRA) of P rank score [48].

## Pairwise and network meta-analysis

For each outcome, frequentist random-effects pairwise and network meta-analyses were performed using the restricted maximum likelihood estimator [44], allowing for heterogeneity in treatment effects between studies. Pairwise analyses were conducted to compare the primary efficacy outcomes of all direct treatment comparisons, using the R metafor package [34]. For two or more studies reporting on the same outcome included in our NMA, direct and indirect (and mixed) comparisons were accomplished through the netmeta package of R [49].

## Heterogeneity measures

We assessed statistical heterogeneity in each pairwise comparison with Q statistic, $\tau^2$ statistic, $I^2$ statistic, and P-value. The categories of heterogeneity we defined were: $I^2 < 30\%$, low; $I^2 = 30–70\%$, moderate; and $I^2 > 70\%$, high. A 2-sided p-value of $< 0.05$ was regarded as statistically significant.

## Certainty of the Evidence

The grading of recommendations assessment, development and evaluation (GRADE) approach for NMA was used to rate the certainty of evidence of NMA estimates [50]. Two investigators rated the certainty of each treatment comparison independently and resolved discrepancies by discussions and, if necessary, consulted with a third party.

### Subgroup and sensitivity analyses

Prespecified subgroup analyses, regression analyses, and sensitivity analyses were performed to determine whether the results were affected by the patient severity, treatment protocol, and follow-up time.

Based on our preliminary meta-regression results, we conducted subgroup analyses only based on pharmacological interventions. The primary outcomes were separately analyzed for mild patients, moderate to severe patients (non-ICU at admission), critically ill patients (ICU), and post-COVID patients as these patients may respond differently to treatments.

Sensitivity analyses were conducted by restricting the analyses to only RCTs, only published studies, and excluding studies with high/serious RoB. Through sensitivity analysis, rstudent value, dffits value, cook.d value, cov.r value, TAU value, QE value, HAT value, and Weight value are used to describe the performance of the NMA model [51].

## Results

### Study overview

The initial search identified 1229 articles. These studies were assessed for inclusion using the prespecified inclusion and exclusion criteria described in the methods. The title and abstract of 610 articles were assessed, and 143 studies were found suitable for full-text review. After excluding 15 studies, 129 studies including 109 RCTs (S2 Table 1) and 20 observation studies (S2 Table 2) were included in qualitative analysis and 58 RCTs were finally included in our NMA (Figure 1). A total of 20,447 COVID-19 patients were included. The characteristics and reference list of included studies are presented in Table 1.

The certainty of evidence ranged from high to very low. Most of the RCTs were at unclear risk of bias (71 of 109), and RCTs at high (19 of 109) or low risk of bias (19 of 109) were both less than one-fifth. Only 17.4% of current evidence on pharmacological management of COVID-19 is supported by moderate or high certainty and can be translated to practice and policy; the remaining 82.6% are of low or very low certainty and warrant further studies to establish firm conclusions. For both pairwise meta-analysis and NMA, the primary and secondary outcomes presented no evidence of significant heterogeneity. The certainty of evidence (GRADE) for each outcome is summarized in Supplementary Appendix 3.

**Pairwise meta-analysis**

Primary outcomes: clinical effective rate

Among 121 adjusted double-armed trials related to COVID-19 symptoms of concern that were included in the pairwise meta-analysis, the summary estimates showed that, compared to control (placebo/SOC), there was a significant reduction in all-cause mortality (OR=–0.26, 95%CI [–0.41, –0.10]), hospital admission (OR=–0.46, 95%CI [–0.83, –0.10]), hospital days (SMD=–0.29, 95%CI [–0.58, –0.01]), ICU admission (OR=–0.52, 95%CI [–1.52, 0.49]), ICU days (SMD=–0.24, 95%CI [–0.46, –0.02]), MV requirement (OR=–0.34, 95%CI [–0.67, –0.01]) or MV days (SMD=–0.71, 95%CI [–1.00, –0.43]).

A total of 47 studies including 45 RCTs and 2 observational trials with 7337 patients reported all-cause mortality. Anti-inflammatory drugs for COVID-19, including tocilizumab (OR=–2.27, 95%CI [–5.43, 0.89]), standard-dose OP-101 (OR=–2.05, 95%CI [–5.16, 1.05]), methylprednisolone pulse therapy (MPPT) (OR=–1.99, 95%CI [–3.56, –0.41]), Xuebijing injection (XBJI) (OR=–1.98, 95%CI [–4.14, 0.18]), high-dose sarilumab (OR=–1.95, 95%CI [–4.94, 1.05]), tranilast (OR=–1.39, 95%CI [–3.64, 0.86]), azithromycin (AZM) (OR=–1.12, 95%CI [–4.34, 2.11]), favipiravir (OR=–0.97, 95%CI [–5.01, 3.07]), renin-angiotensin aldosterone system inhibitors (RAASi) (OR=–0.87, 95%CI [–3.34, 1.60]), zilucoplan (OR=–0.81, 95%CI [–2.14, 0.52]), rivaroxaban (OR=–0.81, 95%CI [–2.01, 0.39]), anakinra (OR=–0.76, 95%CI [–1.55, 0.03]), nano-curcumin (OR=–0.69, 95%CI [–2.04, 0.66]), epoprostenol (OR=–0.69, 95%CI [–1.61, 0.23]), imatinib (OR=–0.63, 95%CI [–1.30, 0.03]), namilumab (OR=–0.53, 95%CI [–1.61, 0.55]), clopidogrel (OR=–0.42, 95%CI [–0.99, 0.15]), metoprolol (OR=–0.41, 95%CI [–3.32, 2.51]), dapagliflozin (OR=–0.28, 95%CI [–0.70, 0.15]), famotidine (OR=–0.18, 95%CI [–0.76, 0.41]), α-1 antitrypsin (Ⅳ AAT) (OR=–0.13, 95%CI [–1.52, 1.27]), atorvastatin (OR=–0.11, 95%CI [–0.44, 0.21]), dexamethasone (OR=–0.09, 95%CI [–0.46, 0.28]) and infliximab (OR=–0.06, 95%CI [–1.47, 1.34]), showed a reduction of all-cause mortality, compared to control. Among them, OP-101, a hydroxyl-polyamidoamine dendrimer-N-acetylcysteine conjugate that specifically targets activated macrophages, improves outcomes of systemic inflammation and neuroinflammation, with high doses (OR=–1.05, 95%CI [–3.52, 1.42]) not as good as standard doses (OR=–2.05, 95%CI [–5.16, 1.05]), but better than low doses (OR=–0.54, 95%CI [–2.56, 1.48]). Meanwhile, sarilumab, an IL-6 inhibitor, worked better at high doses (OR=–1.95, 95%CI [–4.94, 1.05]) than at standard doses (OR=0.34, 95%CI [–1.22, 1.90]). In addition, L-glutamine (OR=–4.31, 95%CI [–7.10, –1.51]), supplementation with vitamins A, B, C, D, and E (OR=–2.20, 95%CI [–5.16, 0.77]), omega-3 fatty acids (OR=–0.43, 95%CI [–2.66, 1.81]) and high-dose zinc gluconate (OR=–0.15, 95%CI [–4.09, 3.79]) may be effective adjuvant nutritional therapy associated with reduced mortality, while patients treated with a combination of high-dose zinc gluconate and vitamin C supplements (OR=1.46, 95%CI [–1.60, 4.52]) or high-dose vitamin C (OR=1.14, 95%CI [–2.09, 4.36]) even had a higher risk of death compared with control patients. Moreover, four adjusted controlled trial with a small sample size associated with mesenchymal stromal cell (MSC) therapy suggested that MSC has shown very promising efficacy in severe COVID-19, with higher survival rates in the MSC cohort compared to matched control patients [52-55]. Hyperbaric oxygen therapy (HBOT) (OR=0.00, 95%CI [–2.84, 2.84]), sofosbuvir-velpatasvir, Lianhuaqingwen capsule (LHQWC) (OR=0.00, 95%CI [–2.78, 2.78]), QingfeiPaidu capsule (QFPDC) (OR=0.00, 95%CI [–2.78, 2.78]), CytoSorb (OR=0.00, 95%CI [–0.01, 1.33]) was not observed to be associated with reduced mortality. Anakinra (OR=0.66, 95%CI [–0.01, 1.33]), bamlanivimab (OR=0.11, 95%CI [–0.73, 0.94]) and alteplase plus heparin (OR=0.12, 95%CI [–0.98, 1.22]) was at risk of increased mortality.

In terms of hospitalization, there were 10 studies including 1341 patients reporting hospital admission and another 10 studies including 2042 patients reporting hospital days. For outpatients with mild COVID-19, those treated with sofosbuvir/daclatasvir (antiviral drugs for hepatitis C as HCV NS5A inhibitors, OR=-1.35, 95%CI [–3.60, 0.90]), and resveratrol (an antioxidant, OR=–1.12, 95%CI [–3.41, 1.18]) had a lower incidence of hospitalization compared to control patients. In addition, outpatients who received supplementation with vitamins A, B, C, D, and E (OR=–1.39, 95%CI [–2.59, –0.18]) and high-dose vitamin C (OR=–0.36, 95%CI [–2.20, 1.47]) were observed to have fewer hospitalizations, while those received high-dose zinc gluconate (OR=0.36, 95%CI [–1.12, 1.84]) or high-dose zinc gluconate plus vitamin C (OR=0.70, 95%CI [–0.70, 2.10]) were observed even more hospitalizations. Among hospitalized patients with moderate-severe COVID-19, more patients were discharged during the study period in the treatment groups of nanomumab (OR=–0.58, 95%CI [–1.38, 0.22]), anakinra (OR=–0.49, 95%CI [–1.24, 0.25]), infliximab (OR=–0.38, 95%CI [–1.44, 0.68]), and mesenchymal stem cell therapy (MSCT) (OR=–0.35, 95%CI [–1.43, 0.73]) than in the control group. Moreover, mesenchymal stem cell therapy (MSCT) (SMD=–0.76, 95%CI [–1.45, –0.07]), L-glutamine (SMD=–0.75, 95%CI [–0.94, –0.56]), namilumab (SMD=–0.65, 95%CI [–1.04, –0.27]), α-1 antitrypsin (Ⅳ AAT) (SMD=–0.64, 95%CI [–1.37, 0.08]), CytoSorb (SMD=–0.62, 95%CI [–1.19, –0.04]), azithromycin (AZM) (SMD=–0.46, 95%CI [–0.84, –0.08]), anakinra (SMD=–0.17, 95%CI [–0.34, 0.01]) were observed to be effective in shortening the length of hospital stay. On the contrary, the treatment groups of infliximab (SMD=0.57, 95%CI [0.06, 1.07]) and imatinib (SMD=0.18, 95%CI [–0.02, 0.38]) showed longer hospital stays than the control group. Discontinuation of renin-angiotensin aldosterone system inhibitors (RAASi) (SMD=0.12, 95%CI [–0.15, 0.40]) was associated with a lower hospitalization length than continuation of RAASi within 30 days [56].

In total, 9 studies with 1021 patients reported ICU admission and 14 studies with 2342 patients reported ICU days or ICU-free days during the study period. Among the non-hospitalized patients, none of those treated with L-glutamine (OR=–5.47, 95%CI [–8.26, –2.69]) were hospitalized compared to the control group with a 53.48% hospitalization rate. For hospitalized patients outside the ICU, azithromycin (AZM) (OR=–1.27, 95%CI [–2.89, –0.34]) were associated with reduced ICU occupancy of patients. While, two studies suggested that either 7-day (OR=–1.27, 95%CI [–3.61, 1.06]) or 30-day (OR=–0.07, 95%CI [–0.76, 0.63]) discontinuation of renin-angiotensin aldosterone system inhibitors (RAASi) increased the risk of admission to the ICU for hospitalized patients. Additionally, it was suggested that propolis is a natural product with considerable evidence of immunoregulatory and anti-inflammatory activities potential against viral targets, and thus, an open-label RCT used Brazilian green propolis (EPP-AF®) (high-dose: OR=1.01, 95%CI [–0.69, 2.71]); standard-dose: OR=1.74, 95%CI [0.16, 3.33]) as an adjunct therapy for hospitalized COVID-19 patients. However, the results showed that propolis was associated with an increased risk of ICU admission. For ICU-admitted patients, infliximab (OR=–0.81, 95%CI [–2.38, 0.76]) and namilumab (OR=–0.74, 95%CI [–1.78, 0.30]) increased the likelihood of patients being discharged from ICU, while CytoSorb (OR=–0.07, 95%CI [–0.79, 0.92]) showed the risk of reducing the number of discharged patients. Imatinib (SMD=–0.83, 95%CI [–1.04, –0.62]), Xuebijing injection (XBJI) (SMD=–0.75, 95%CI [–1.29, –0.21]), namilumab (SMD=–0.61, 95%CI [–1.23, 0.00]), α-1 antitrypsin (Ⅳ AAT) (SMD=–0.59, 95%CI [–1.31, –0.13]), anakinra (SMD=–0.25, 95%CI [–0.43, –0.08]), sofosbuvir/velpatasvir (SMD=–0.22, 95%CI [–0.66, 0.22]), alteplase plus heparin (SMD=–0.13, 95%CI [–0.68, 0.43]; 2 RCTs, 50 patients), and dexamethasone (SMD=–0.04, 95%CI [–0.27, 0.19]) reduced the length of ICU stay of hospitalized patients. Atorvastatin (SMD=0.00, 95%CI [–0.16, 0.16]) showed no difference in reducing the ICU days compared with control. CytoSorb (SMD=0.17, 95%CI [–0.39, 0.73]), azithromycin (AZM) (SMD=0.27, 95%CI [–0.10, 0.64]), mesenchymal stem cell therapy (MSCT) (SMD=0.30, 95%CI [–0.51, 1.10]), infliximab (SMD=0.31, 95%CI [–0.58, 1.20]) did not reduce ICU days and was at risk of increasing.

Overall, 12 studies with 1612 patients reported the number of patients requiring mechanical ventilation, while another 6 studies with 821 patients reported MV days or MV-free days during the study period. Anakinra (OR=–1.19, 95%CI [–1.58, –0.80]; 2 RCTs, 854 patients), hyperbaric oxygen therapy (HBOT) (OR=–1.10, 95%CI [–3.45, 1.25]), sofosbuvir/velpatasvir (OR=–1.10, 95%CI [–3.40, 1.21]), standard-dose sarilumab (OR=–0.24, 95%CI [–2.38, 0.76]), mesenchymal stem cell therapy (MSCT) (OR=–0.07, 95%CI [–1.13, 1.00]; 2 RCTs, 73 patients), standard-dose propolis (OR=–0.03, 95%CI [–0.66, 0.60]). Either 7-day (OR=–0.87, 95%CI [–3.34, 1.60]) or 30-day (OR=–0.81, 95%CI [–2.38, 0.76]) discontinuation of renin-angiotensin aldosterone system inhibitors (RAASi) increased the demand for MV. Also, the treatment group of high-dose propolis (OR=0.00, 95%CI [–0.62, 0.62]) was no different from control group in MV requirement. Meanwhile, high-dose sarilumab group (OR=0.41, 95%CI [–0.94, 1.75]) even needed more MV support than its control group. Dexamethasone (SMD=–1.12, 95%CI [–1.36, –0.87]), metoprolol (SMD=–0.62, 95%CI [–1.54, 0.29]), mesenchymal stem cell therapy (MSCT) (SMD=–0.59, 95%CI [–1.40, 0.22]), CytoSorb (SMD=–0.57, 95%CI [–1.14, 0.00]), imatinib (SMD=–0.56, 95%CI [–0.76, –0.35]), α-1 antitrypsin (Ⅳ AAT) (SMD=–0.45, 95%CI [–1.17, 0.27]) can effectively shorten MV days.

Secondary outcomes

Regarding secondary outcomes, the pairwise meta-analysis, including 16 studies with 2673 patients, showed that anakinra (OR=0.64, 95%CI [0.29, 1.00]), ivermectin mucoadhesive nanosuspension nasal spray (Ivermectin MMNS) (OR=0.23, 95%CI [–0.32, 0.77]), mesenchymal stem cell therapy (MSCT) (OR=0.15, 95%CI [–0.81, 1.11]), Lianhuaqingwen capsule (LHQWC) (OR=0.11, 95%CI [–0.24, 0.45]), sofosbuvir/velpatasvir (OR=0.09, 95%CI [–0.65, 0.83]) and famotidine (OR=0.07, 95%CI [–0.35, 0.49]) were better than control in terms of overall symptom improvement. Tetra sodium pyrophosphate (TSPP) (anosmia: OR=3.97, 95%CI [1.13, 6.81]), hydrogen peroxide (H2O2) (dysgeusia: OR=0.92, 95%CI [–0.86, 2.70]; hyposmia: OR=0.07, 95%CI [–0.77, 1.86]), mometasone furoate nasal spray (MFNS) (anosmia: OR=0.22, 95%CI [–0.28, 0.72]; 3 RCTs, 183 patients) and ivermectin mucoadhesive nanosuspension nasal spray (Ivermectin MMNS) (anosmia: OR=0.07, 95%CI [–0.51, 0.65]) improved chemosensory dysfunctions better than control. Melatonin (OR=0.45, 95%CI [–0.64, 1.54]), Lianhuaqingwen capsule (LHQWC) (OR=0.02, 95%CI [–0.36, 0.40]) and QingfeiPaidu capsule (QFPDC) (OR=0.01, 95%CI [–0.37, 0.39]) showed better results in improving fatigue than the control group. Shumian capsule (SMC) showed significant effect on improving depression (OR=1.59, 95%CI [–1.46, 4.65]) and insomnia (OR=0.99, 95%CI [–2.22, 4.20]), but no effect on improving anxiety (OR=0.04, 95%CI [–1.46, 1.38]). Lianhuaqingwen capsule (LHQWC) (SMD=–2.02, 95%CI [–2.31, –1.73]), methylprednisolone pulse therapy (MPPT) (SMD=–1.54, 95%CI [–2.11, –0.97]), famotidine (SMD=–0.88, 95%CI [–1.16, –0.59]), mesenchymal stem cell therapy (MSCT) (SMD=–0.63, 95%CI [–1.32, –0.05]), high-dose zinc gluconate plus vitamin C (SMD=–0.31, 95%CI [–0.69, 0.07]), high-dose vitamin C (SMD=–0.29, 95%CI [–0.69, 0.11]), sofosbuvir/velpatasvir (SMD=–0.23, 95%CI [–0.67, 0.21]) and high-dose zinc gluconate (SMD=–0.17, 95%CI [–0.55, 0.21]) compared to control had a reduction in the progression of COVID-19 disease. However, tocilizumab (SMD=0.15, 95%CI [–0.34, 0.64]) and Shufeng Jiedu capsules (SFJDC) (SMD=0.49, 95%CI [0.03, 0.95]) had no positive effect on recovery duration. Ivermectin mucoadhesive nanosuspension nasal spray (Ivermectin MMNS) (Aref 2021: SMD=–1.82, 95%CI [–2.26, –1.38], 114 patients; Aref 2022: SMD=–0.47, 95%CI [–0.88, –0.06], 96 patients) was significantly effective in shortening olfactory recovery time, while patients treated with intranasal betamethasone sodium phosphate drops (IBSPD) (SMD=0.00, 95%CI [–0.24, 0.24]) and mometasone furoate nasal spray (MFNS) (SMD=0.04, 95%CI [–0.35, 0.43]) was at risk of prolonged olfactory recovery duration. Xuebijing injection (XBJI) (SMD=–1.97, 95%CI [–2.60, –1.34]), Lianhuaqingwen capsule (LHQWC) (SMD=–1.28, 95%CI [–1.53, –1.02]) and Shufeng Jiedu capsules (SFJDC) (SMD=–0.73, 95%CI [–1.20, –0.26]) compared to control reduced fatigue recovery duration. Hydrogen (H2) show superiority over the control group in reducing the duration of fatigue (SMD=–0.02, 95%CI [–0.58, 0.53]) or insomnia (SMD=0.00, 95%CI [–0.55, 0.55]) symptoms. Detailed pairwise meta-analysis results of secondary outcomes are presented in Supplementary Appendix 5.

## Network meta-analysis

We visualized the network of comparisons as shown in Figure 3. In the network, each regimen is represented by a unique node, meaning different nodes were designated for different interventions or different dosages of the same drug. Lines indicate direct head-to-head comparison of agents, and the thickness of the line corresponds to the number of trials in the comparison. The size of the node corresponds to the number of total patients behind the intervention. Figure 4 is the league table of treatment rankings based on cumulative probability plots and SUCRAs of all-cause mortality, AE, and SAE, presented as two dimensions of efficacy and safety outcomes. The global inconsistency was not significant for all the outcomes considered. Tests of local inconsistency did not show any inconsistent loops (see also Supplementary Appendix 4).

The results of the network meta-analysis (NMA) are presented in Figure 3 for the primary outcomes. In terms of all-cause mortality, we evaluated 41 studies.

When compared to control (placebo/SOC), L-glutamine (OR=0.01, 95%CI [0.00, 0.29]), favipiravir (OR=0.10, 95%CI [0.00, 2.47]), supplementation with vitamins A, B, C, D, and E (OR=0.11, 95%CI [0.00, 2.59]), standard-dose OP-101 (OR=0.13, 95%CI [0.01, 2.60]), methylprednisolone pulse therapy (MPPT) (OR=0.14, 95%CI [0.02, 0.92]), Xuebijing injection (XBJI) (OR=0.14, 95%CI [0.01, 1.52]), standard-dose sarilumab (OR=0.14, 95%CI [0.01, 3.49]), tranilast (OR=0.25, 95%CI [0.02, 3.00]), mesenchymal stem cell therapy (MSCT) (OR=0.25, 95%CI [0.03, 2.24]) significantly reduced the mortality rate. Since the NMA results of these primary efficacy outcomes were consistent with previous pairwise meta-analysis results, further confirming the confidence of the results, detailed information of studies, included in the analysis for all-cause mortality, hospital admission, hospital days, ICU admission, ICU days, MV requirement and MV days, are presented in Supplementary Appendix 4.

Regarding safety, we evaluated 19 studies in terms of adverse events (AEs) and 10 studies in terms of serious adverse events (SAEs). No significant differences were found between the included compounds. A number of pharmacological interventions were worse than control in terms of AEs, including high-dose zinc (OR=21.87, 95%CI [1.25, 383.45]), high-dose vitamin C (OR=56.11, 95%CI [3.26, 966.29]), the combination of high-dose zinc gluconate and vitamin C supplements (OR=42.59, 95%CI [2.49, 729.67]), tocilizumab (OR=9.64, 95%CI [2.76, 33.75]), Shugan Jieyu capsule (SGJYC) (OR=4.04, 95%CI [0.44, 36.83]), Bufei Huoxue capsule (BFHXC) (OR=2.14, 95%CI [0.51, 8.95]), the combination of QingfeiPaidu capsule (QFPDC) and Lianhuaqingwen capsule (LHQWC) (OR=1.48, 95%CI [0.72, 3.03]), standard-dose sarilumab (OR=1.08, 95%CI [0.40, 2.90]), zilucoplan (OR=1.07, 95%CI [0.37, 3.10]) and anakinra (OR=0.95, 95%CI [0.67, 1.33]). Dexamethasone (OR=0.19, 95%CI [0.07, 0.53]), standard-dose (OR=0.22, 95%CI [0.04, 1.13]) or high-dose (OR=0.33, 95%CI [0.08, 1.33]) of propolis, QingfeiPaidu capsule (QFPDC) (OR=0.66, 95%CI [0.32, 1.35]), Lianhuaqingwen capsule (LHQWC) (OR=0.71, 95%CI [0.45, 1.14]), Persian medicine herbal formulations (PMHF) (OR=0.88, 95%CI [0.58, 1.33]), high-dose sarilumab (OR=0.88, 95%CI [0.32, 2.39]) proved to reduce AEs when compared to control. In terms of SAEs, therapeutic-dose apixaban (OR=1.84, 95%CI [0.53, 6.47]) and prophylactic-dose apixaban (OR=1.16, 95%CI [0.31, 4.38]) were at higher risk than control. Besides, high-dose (OR=0.25, 95%CI [0.05, 1.22]) or standard-dose (OR=0.36, 95%CI [0.08, 1.59]) of propolis, alteplase plus heparin (OR=0.26, 95%CI [0.06, 1.09]), tocilizumab (OR=0.29, 95%CI [0.01, 8.31]), anakinra (OR=0.50, 95%CI [0.25, 0.97]), dexamethasone (OR=0.53, 95%CI [0.13, 2.12]), zilucoplan (OR=0.57, 95%CI [0.13, 2.56]), imatinib (OR=0.67, 95%CI [0.26, 1.77]), dapagliflozin (OR=0.77, 95%CI [0.32, 1.88]), aspirin (OR=0.80, 95%CI [0.20, 3.20]) proved to be better in terms of SAEs when compared to control.

**Main clinical symptoms**

Among 109 RCTs and 20 confounding-adjusted observational studies including a total of 20,447 patients, we collected treatment strategies for five major types of COVID-19 sequelae of concern including olfactory dysfunction, cardiovascular symptoms, organ failure, mental disorders and inflammatory response. For each outcome, we conducted appropriate qualitative and quantitative NMA analyses.

Olfactory dysfunction

13 studies with 937 patients reporting improvement of smell scores including TDI score, UPSIT score, Iran-SIT score or VAS score in the treatment of long-COVID symptoms of anosmia, hyposmia or dysgeusia were included in our NMA (see Supplementary Appendix 6). Saline nasal irrigation (SNI) (SMD=21.10, 95%CI [16.91, 25.30]), nitrilotriacetic acid trisodium (NAT) (SMD=7.40, 95%CI [5.79, 9.01]), tetra sodium pyrophosphate (TSPP) (SMD=3.69, 95%CI [2.61, 4.77]), sodium gluconate (SMD=3.01, 95%CI [1.92, 4.09]), mometasone furoate nasal spray (MFNS) (SMD=0.80, 95%CI [–0.08, 1.67]) and olfactory training (OT) (SMD=0.24, 95%CI [–0.64, 1.12]) compared to control showed significant improvement in the olfactory score in post-COVID patients in analysis of RCTs. Also, saline nasal irrigation (SNI) (SMD=20.87, 95%CI [16.58, 25.15]), nitrilotriacetic acid trisodium (NAT) (SMD=7.16, 95%CI [5.32, 8.99]), tetra sodium pyrophosphate (TSPP) (SMD=3.45, 95%CI [2.06, 4.84]), sodium gluconate (SMD=2.77, 95%CI [1.38, 4.16]), palmitoylethanolamide and luteolin combined with olfactory training (OT) (SMD=0.58, 95%CI [0.03, 1.13]), mometasone furoate nasal spray (MFNS) (SMD=0.56, 95%CI [–0.68, 1.79]) and omega-3 fatty acids plus olfactory training (OT) (SMD=0.08, 95%CI [–0.81, 0.97]) proved to improve anosmia better when compared to olfactory training (OT). However, mometasone furoate nasal spray (MFNS) plus olfactory training (OT) (SMD=–0.12, 95%CI [–0.71, 0.47]) and advanced olfactory training (AOT) (SMD=0.26, 95%CI [–1.12, 0.60]) showed even less improvement of olfactory symptom than olfactory training (OT).

Cardiovascular symptoms

A total of 32 RCTs including 7336 patients with COVID-19 were included in the NMA of treatment for cardiovascular symptoms characterized by thrombosis events. Potential COVID-19-related cardiovascular drugs for the treatment of sequelae thrombosis can be divided into two main categories: anti-platelet aggregation drugs like aspirin and clopidogrel, and anticoagulants for thrombosis prophylaxis.

Certain different types of anticoagulants based on different therapeutic mechanisms to prevent blood clots in hospitalized patients with critically ill hospitalized [57-59], noncritically ill hospitalized patients or outpatients [60, 61] infected with COVID-19 are mainly as follows:

1) Vitamin K antagonist. By antagonizing vitamin K, the liver synthesis of prothrombin and factors VII, IX and X are reduced and anticoagulates. For instance, warfarin is a typical type of oral anticoagulant.

2) Indirect thrombin inhibitors. Through the interaction with antithrombin (AT-III), the activities of factors X a and IIA are indirectly inhibited to play an anticoagulant role. Typical drugs are injected anticoagulants including heparin and low-molecular-weight heparin characterized by enoxaparin, bemiparin and dalteparin.

3) Direct thrombin inhibitors. By inhibiting thrombin, they prevent fibrinogen lysis to fibrin, and block the final steps of the clotting cascade and clots forming. Among them, monovalent thrombin inhibitors (Dabigatrun ester, agattriban, etc.) can directly inhibit thrombin; bivalent thrombin inhibitors (bivaludin, recombinant hirudin, etc.) can directly inhibit thrombin, and also separate thrombin and fibrin to achieve anticoagulant effect. They are suitable for stroke prevention of atrial fibrillation, treatment and prevention of acute coronary syndrome and VTE (including primary and secondary).

4) Factor Xa Inhibitors. Activation factor Xa is a kind of serine protease, in the coagulation cascade in the position of external and external sources of coagulation intersection, factor Xa inhibitors make the endogenous and exogenous pathway of the coagulation waterfall interrupted, according to whether dependent on AT-III factor can be divided into indirect and direct inhibitors. Indirect factor Xa inhibitors require AT-III factor as a cofactor, and cannot inhibit factor Xa bound by prothrombin complex. Direct factor Xa inhibitors act directly on the active center of factor Xa molecules, inhibiting both the free factor Xa in plasma and the factor Xa bound by prothrombin complex to play an anticoagulant role. Rivaroxaban and apixaban are both oral anticoagulant of this type.

There are also other types of anticoagulant treatments on trial such as atorvastatin (an antilipidemic drug), dapagliflozin (an oral hypoglycemic agent), renin-angiotensin-aldosterone system inhibitors (RAASi) and CytoSorb adsorber (CytoSorbents, Monmouth Junction, NJ) [62].

To evaluate the efficacy of different anticoagulant drugs on cardiovascular symptoms associated with COVID-19, we included 5 studies with 2494 patients in the NMA for arterial thrombotic events (ATEs) or major thrombotic events (MTEs) and 5 studies with 1953 patients for venous thrombotic events (VTEs) or major thrombotic events (MTEs) (see Figure 6). Rivaroxaban proved to reduce both ATEs (OR=0.33, 95%CI [0.01, 8.19]) and VTEs (OR=0.12, 95%CI [0.01, 0.97]). Also, dapagliflozin (OR=0.79, 95%CI [0.53, 1.19]) was associated with reduced thrombotic events compared to the control. Atorvastatin (OR=0.68, 95%CI [0.24, 1.92]) proved to be effective in reducing VTEs. Continuation of RAASi (OR=0.04, 95%CI [0.00, 0.33]) reduced ATEs compared to discontinuation of RAASi within 7 days. In terms of renin-angiotensin-aldosterone system inhibitor management, randomized controlled trials indicated that RAASi continuation in participants hospitalized with COVID-19 appeared safe. Although discontinuation of RAS-inhibition in COVID-19 had no significant effect on the maximum severity of COVID-19, discontinuation increased brain natriuretic peptide levels and may increase the risk of acute heart failure, and thus where possible, RAASi should be continued [56, 63]. As an antiplatelet aggregator, clopidogrel compared to control showed no effect on reducing arterial clots (OR=2.82, 95%CI [0.11, 69.78]) but may be associated with reduced venous clots (OR=0.72, 95%CI [0.36, 1.44]). While, as a type of blood purification technology potential to treat cytokine storm and lethal inflammation in critically ill and cardiac surgery patients associated with COVID-19 [62], CytoSorb had a greater risk of increasing blood clots (OR=1.52, 95%CI [0.49, 4.69]) in a pilot trial [64].

In terms of drug safety, 6 studies with 1458 patients reported bleeding events (BEs) or major bleeding events (MBEs). Therapeutic (OR=4.43, 95%CI [1.23, 15.92]) and prophylactic (OR=3.17, 95%CI [0.84, 11.96]) doses of apixaban, atorvastatin (OR=2.30, 95%CI [0.79, 6.71]) and aspirin (OR=1.93, 95%CI [0.47, 7.87]) were observed to have a greater risk of major bleeding, while alteplase plus heparin (OR=1.02, 95%CI [0.19, 5.43]) and clopidogrel (OR=0.93, 95%CI [0.13, 6.71]) showed no difference compared to control in terms of MBEs.

The anticoagulants used in the included studies were divided into four categories (see Supplementary Appendix 5): no treatment, prophylactic dose, intermediate dose and therapeutic dose. However, the definition of intermediate dose in some studies was included in that of the prophylactic dose. To perform NMA, the intermediate-dose group was included in the prophylactic dose group; the anticoagulants were generally classified into three groups (no treatment, prophylactic dose and treatment dose).

In the NMA of different doses of anticoagulants, 7 studies with 4506 patients reported ATEs, 8 studies with 4504 patients reported VTEs and 6 studies with 4784 patients reported MBEs. The results showed that the therapeutic (OR=0.19, 95%CI [0.01, 5.63]) dose was associated with a lower risk of ATEs than prophylactic or intermediate dose (OR=0.33, 95%CI [0.01, 8.19]) or control (see Figure 7A). The therapeutic (OR=0.04, 95%CI [0.01, 0.38]) dose were also associated with a lower risk of VTEs than prophylactic or intermediate dose (OR=0.12, 95%CI [0.01, 0.97]) (see Figure 7B). Additionally, in terms of MBEs, the treatment dose (OR=1.86, 95%CI [1.19, 2.89]) was associated with an increased risk of major bleeding compared to that observed with the prophylactic dose (see Figure 7C).

Heparin is mainly used in anticoagulation therapy and assumed to have anticoagulant effects as well as anti-inflammatory and antiviral effects through neutralization of ribonucleic acid histones and cytokines [65]. Many studies have reported the efficacy of prophylactic and therapeutic doses of anticoagulants in patients with COVID-19 [66-68], but only a few high-quality randomized controlled trials (RCTs) have been conducted [59, 60, 69-71]. Several trials suggested benefit from therapeutic dose anticoagulation (heparin or LMWH) compared to usual care pharmacologic thromboprophylaxis in patients hospitalized with moderate COVID-19 but potential harm in patients with more severe COVID-19 who required ICU-level care, raising concerns about the risk of bleeding with therapeutic dose anticoagulation [72].

## Organ failure

We included 25 studies including 4187 patients to assess the treatments for COVID-19-induced organ failure. For moderate-severe hospitalized patients, α-1 antitrypsin (IV AAT), alteplase plus heparin (2 studies), anakinra (3 studies), Bufei Huoxue capsule (BFHXC), dapagliflozin, dexamethasone, imatinib, metoprolol, methylprednisolone pulse therapy (MPPT), renin-angiotensin-aldosterone system inhibitors (RAASi), tocilizumab, sofosbuvir/velpatasvir, sarilumab, hyperbaric oxygen therapy (HBOT) and mesenchymal stem cell therapy (MSCT) (4 studies) are potential tolerated treatments for multiple organ failure typical of dyspnea, hypoxia, respiratory failure, and even ARDS.

Patients with severe COVID-19 typically develop a febrile pro-inflammatory cytokinemia with accelerated progression to acute respiratory distress syndrome (ARDS) [73-75], a pathological entity characterized by alveolar epithelial and lung endothelial injury, excessive protease activity, and dysregulated airway inflammation [76, 77] that is associated with prolonged invasive mechanical ventilation, prolonged hospitalization, and long-term disability [78]. Treatments reported in RCTs for ARDS patients with COVID-19 include α-1 antitrypsin (IV AAT), dexamethasone, imatinib, metoprolol, methylprednisolone pulse therapy (MPPT), and mesenchymal stem cell therapy (MSCT), which may be effective in improvement of oxygen saturation or sepsis-related organ failure (SOFA) score. Among them, intravenous dexamethasone plus standard care was associated with a statistically significant increase in days alive and ventilator-free over 28 days compared to standard care alone [79]. While imatinib did not show an effect in reducing the time to discontinuation of ventilation and supplemental oxygen for more than 48 consecutive hours in a double-blind RCT, the observed effects on survival and duration of mechanical ventilation suggested that it may provide clinical benefit to hospitalized COVID-19 patients, which still requires to be validated by further studies [80]. In a pilot trial, intravenous metoprolol administration to patients with COVID-19-associated ARDS proved to be safe and effective in reducing exacerbated lung inflammation and improving oxygenation. Also, methylprednisolone pulse therapy (MPPT) was suggested in a single-blind RCT to be an efficient therapeutic agent for hospitalized severe COVID-19 patients at the pulmonary phase [81]. In addition, four observational clinical trials with small samples of mesenchymal stem cell (MSCs) therapy for severe or critical COVID-19 patients suggested that MSCs are safe and can significantly improve respiratory distress and reduce inflammatory biomarkers in some critically ill COVID-19-induced ARDS cases. In addition, four observational clinical trials with small samples of mesenchymal stem cell (MSCs) therapy for severe or critical COVID-19 patients suggested that MSCs are safe and can rapidly improve respiratory distress and decrease inflammatory cytokine levels, potential to be effective in reducing mortality rates of severe COVID-19-induced ARDS [54, 55, 82] and even pulmonary fibrosis [83]. However, these observations need to be confirmed in large randomized multicenter clinical trials designed to demonstrate efficacy.

In addition to those for ARDS, several other drugs have been reported to be associated with COVID-19-related respiratory failure and hypoxemia. A phase 1/2 clinical study reported the combination of alteplase (tPA) bolus plus heparin is safe in severe COVID-19 respiratory failure, while a phase 3 study is warranted given the improvements in oxygenation and promising observations in ventilator-free days (VFD) and mortality [84]. Besides, anakinra has already proved in a double-blind randomized controlled phase 3 trial to have significant effects on decreased severe respiratory failure (SRF) and restored pro-/anti-inflammatory balance [85]. In terms of open trials of treatments for COVID-19 hyoxemia, tocilizumab was reported to be effective in improving hypoxia without unacceptable side effects and had a significant impact on the negative conversion time of viral infection [86]. Hyperbaric oxygen therapy (HBOT) was also supportive as an adjuvant treatment for patients with COVID-19 severe hypoxemia [87]. In addition, a double-blind RCT in Brazil reported the significant efficacy of hydrogen peroxide (H2O2) as an auxiliary treatment for COVID-19-related cough, loss of taste, and hyposmia [88]. While, prostacyclin infusion compared to placebo, resulted in a measurable decrease in endothelial glycocalyx shedding (syndecan-1) at 24 h, suggesting a protective effect on the endothelium, which may be related to the observed reduction in organ failure [89].

Also, for post-COVID patients with symptoms of dyspnea and decreased physical and mental function, a single-blind RCT reported an online breathing and wellbeing program (ENO Breathe) for people with persistent symptoms following COVID-19, which improves the mental component of HRQoL and elements of breathlessness in people with persisting symptoms after COVID-19 [90]. It was suggested that mind-body and music-based approaches, including practical, enjoyable, symptom-management techniques might have a role in supporting recovery.

Mental disorders

We reviewed 34 studies with a total of 3,809 patients to evaluate the treatment of mental disorders related to the sequelae of COVID-19, including fatigue, anxiety, stress, depression, insomnia, cognitive impairment, functional impairment, post-traumatic stress symptoms (PTSS), post-traumatic stress disorder (PTSD) and so on.

In terms of Chinese patent medicine, Hua Shi Bai Du granule (Q-14) [91], Lianhuaqingwen capsule (LHQWC) [92], Shufeng Jiedu capsule (SFJDC) [93], Shugan Jieyu capsule (SGJYC) [94] and Shumian capsule (SMC) [95] have been reported in RCTs as potential drugs to improve COVID-19 symptoms such as fatigue, anxiety, stress, depression, insomnia, etc. Persian medicine herbal formulations (capsules and decoction) have also been reported that it significantly reduced symptoms such as dry cough, dyspnea, muscle pain, headache, fatigue, anorexia, chills, runny nose, phlegm and dizziness [96]. In a double-blind RCT for COVID-19 outpatients, sofosbuvir/daclatasvir significantly reduced the number of patients with fatigue and dyspnoea after 1 month [97]. Vitamin A supplementation also demonstrated efficacy in improving some clinical and paraclinical symptoms in patients with COVID-19 [98]. However, future studies should evaluate efficacy with a larger sample size and compare different dosages. Anhydrous Enol-Oxaloacetate (AEO) is a nutritional supplement, has been anecdotally reported to relieve physical and mental fatigue, but further study of oxaloacetate supplementation for the treatment of long COVID is warranted [99].

Among randomized controlled trials of non-drug treatments for psychological symptoms typical of fatigue and anxiety, aromatherapy [100, 101], cognitive behavioral therapy (CBT) [102-104], narrative exposure therapy (NET) [105], neurofeedback therapy (NFBT) [106], guided imagery (GI) [107] and mandala colouring (MC) [108] were reported to be potentially effective treatments for long-term symptoms in post-COVID-19 patients. RCTs dealing with sequelae of COVID-19 also refer to treatments such as inspiratory muscle training (IMT), manual diaphragm release technique combined with inspiratory muscle training, respiratory (inspiratory/expiratory) muscle training (RMT), progressive muscle relaxation (PMR), aerobic training and yoga. However, the implementation of these treatments is difficult to quantify and validate, so further research into related mechanisms and more large-sample trials are still required. A single-blind RCT proved that H2 inhalation might represent a safe, effective approach for accelerating early function restoration in post-COVID-19 patients [109]. In addition, another double-blind RCT indicated that hyperbaric oxygen therapy (HBOT) improves disruptions in white matter tracts and alters the functional connectivity organization of neural pathways attributed to cognitive and emotional recovery in post-COVID-19 patients [110].

Inflammatory response

It was reported that IL-1β, IL-6, IL-8, and sTNFR1 were all increased in COVID-19 patients. Neutrophils undergo immunometabolic reprogramming in severe COVID-19 illness, associated with lower IL-10, increased cytosolic PKM2 (pyruvate kinase M2), phosphorylated PKM2, HIF-1α (hypoxia-inducible factor-1α) and lactate. Also, the production and sialylation of AAT increased in COVID-19 as an anti-inflammatory response but was overwhelmed in severe illness, with the IL-6:AAT ratio markedly higher in patients requiring ICU admission. In critically ill patients with COVID-19, increased IL-6:AAT predicted prolonged ICU stay and mortality and provided potential implications for clinical treatment. [75] Recently, a systematic review and meta-analysis reported over 20 inflammatory and vascular biomarkers in post-COVID-19 syndrome: higher levels of C-reactive protein, D-dimer, lactate dehydrogenase, and leukocytes were found in COVID-19 survivors with PCS than in those without PCS; lymphocytes and interleukin-6 were also significantly higher in PCS than non-PCS cases; while no significant differences were noted in ferritin, platelets, troponin, and fibrinogen, etc. [27] In addition, multiple recent genetic association studies have reported an association between C-C chemokine receptor 5 (CCR5) and the severity of COVID-19 [111-113]; thus, this receptor could also be involved in the pathogenesis of long COVID.

In our evaluation, we included 45 studies with 7113 patients associated with potential treatment strategies targeted at COVID-19-related immune dysfunction and inflammatory response. Alpha-1 antitrypsin (AAT) is a typical serine protease inhibitor [114-116] that protects the airway from neutrophil elastase (NE) damage. It is a potent anti-inflammatory and immunomodulator that regulates the production and activity of several key pro-inflammatory cytokines, including IL-6, IL-1β, IL-8, and TNF-α [117-121]. It was reported in a double-blind RCT that treatment with IV AAT resulted in decreased inflammation and was safe and well tolerated [122]. Soluble CD24 (CD24Fc), which is linked to the Fc domain of human IgG1, potential to be developed to attenuate inflammation associated with viral infections, autoimmunity, and graft-versus-host diseases [123-125]. It was reported that CD24Fc rapidly down-regulated systemic inflammation and restored immune homeostasis in COVID-19 patients, supporting further development of CD24Fc as a novel drug for the treatment of severe COVID-19 [126]. Combined metabolic activators (CMA), consisting of L-serine (serine), NAC, NR, and L-carnitine tartrate (LCAT, the salt form of L-carnitine), is a potential treatment for nonalcoholic fatty liver disease based on integrative analysis of multi-omics data derived from different metabolic conditions [127-130]. In the phase 2/3 clinical trial, mild-to-moderate COVID-19 patients treated with CMAs showed significant improvement in plasma levels of proteins and metabolites associated with inflammation and antioxidant metabolism as compared to placebo [131]. Type I IFNs-α/β is broad-spectrum antivirals, exhibiting both direct inhibitory effects on viral replication and supporting an immune response to clear virus infection [132], which had shown effects on the accelerated resolution of lung abnormalities associated with SARS [133]. Arbidol (ARB) (Umifenovir) (ethyl-6-bromo-4-[(dimethylamino)methyl]-5-hydroxy-1-methyl-2 [(phenylthio)methyl]-indole-3-carboxylate hydrochloride monohydrate), a broad spectrum direct-acting antiviral, induces IFN production and phagocyte activation, displaying antiviral activity against respiratory viruses, including coronaviruses [134]. One study showed that treatment with IFN-α2b with or without arbidol significantly shortened the duration of elevated blood levels of inflammatory markers IL-6 and CRP as well as the duration of detectable viruses in the upper respiratory tract [135]. Zilucoplan is an investigational macrocyclic peptide inhibitor of the terminal complement protein C5 that prevents both the formation of active C5a and the membrane attack complex C5b-9 and has been clinically tested in neurological disease [136]. However, it did not show a significant effect, but its indicators of respiratory function and clinical results suggest that C5 inhibition may be beneficial, although further large-scale randomized studies are needed [137]. Leronlimab is a CCR5-binding humanized immunoglobulin G4 monoclonal antibody that has been tested in extensive human trials for the treatment of human immunodeficiency virus type 1 infection [138-141] and has been suggested to improve lymphopenia, particularly CD8 T-cell levels, by resolving inappropriate inflammation in acute severe COVID-19 [142]. In an exploratory trial of the treatment of long COVID with CCR5 in combination with the antibody Lehrenzumab, a significant increase in CCR5 on the surface of blood cells in treated symptomatic responders was observed as compared to non-responders or placebo-treated participants [143]. Also, OP-101, as a hydroxyl-polyamidoamine dendrimer-N-acetylcysteine conjugate that specifically targets activated macrophages, improved outcomes of systemic inflammation and neuroinflammation in a phase2a clinical trial for severe COVID-19 [144]. In addition, CIGB-258 is an altered peptide ligand (APL) derived from the cellular stress protein 60 (HSP60), which developed anti-inflammatory effects and increased regulatory T cell (Treg) activity in preclinical models. A study of COVID-19 patients treated with CIGB-258 found that biomarker levels associated with hyperinflammation, such as interleukin (IL)-6, IL-10, tumor necrosis factor (TNF-α), granzyme B, and perforin, significantly decreased during treatment, and Tregs were induced in all patients studied [145]. Moreover, curcumin [146], as an anti-inflammatory herbal-based agent, has been available in different forms, including powder, capsules, and tablets. To facilitate the application of curcumin and improve its stability and solubility, researchers have formulated curcumin with the aid of nanotechnology. In trials with COVID-19 patients, nano-curcumin has shown the potential to regulate increased rates of mRNA expression and cytokine secretion of inflammatory cytokines, especially IL-1β and IL-6, thereby improving clinical manifestation and overall recovery [147, 148]. In addition, photobiomodulation (PBM), also known as low-level light therapy (LLLT), proved to improve the response of serum cytokines including IL-6, IL-8, and TNF-α in mild to moderate COVID-19 [149].

What's more, many other drugs have been studied in clinical trials, involving their effect on the anti-inflammatory response in COVID-19 patients, include: L-glutamine, supplementation with vitamins A, B, C, D, and E, resveratrol, omega-3 fatty acids, propolis, high-dose zinc gluconate plus vitamin C and other such oral nutritional supplements; the combination of QingfeiPaidu capsule (QFPDC) and Lianhuaqingwen capsule (LHQWC), QFPDC, LHQWC, Xuebijing injection (XBJI), Yindan Jiedu granules (YDJDG) and other such Chinese patent medicine; the combination of favipiravir and tocilizumab, favipiravir, tocilizumab, infliximab, namilumab, palmidrol, sarilumab, tranilast, anakinra, baricitinib, nimotuzumab, and other such western medicine.

## Publication bias and sensitivity analysis

The comparison-adjusted funnel plots of the network meta-analysis were suggestive of some publication bias for all-cause mortality (45 RCTs evaluated) (see Supplementary Appendix 4). Few studies reported similar comparisons for hospital admission (8 RCTs evaluated), hospital days (10 RCTs evaluated), ICU admission (8 RCTs evaluated), ICU days (14 RCTs evaluated), MV requirement (11 RCTs evaluated) and MV days (6 RCTs evaluated), which makes difficult the interpretation of the funnel plots for these outcomes (see also Supplementary Appendix 4).

The results of our sensitivity analysis are reported in Supplementary Appendix 4. The assessments of patient severity, treatment protocol, country, and follow-up based on subgroup and regression analysis are also presented in Supplementary Appendix 4.

# Discussion

According to the inclusion and exclusion criteria in the study, a total of 109 randomized controlled trials (RCTs) and 20 confounding-adjusted observational studies included in this study all described the treatment group, control group, and outcome indicators in detail. We analyzed 58 active pharmacologic agents and their combinations in a large-scale analysis incorporating 20,447 COVID-19 patients. We did not limit our inclusions to RCTs and incorporated observational studies of treatments for long COVID. In addition, we included pharmacological interventions for COVID-19 symptoms associated with sensory nerve damage, cardiovascular symptoms, organ failure, psychological disorders and inflammatory response or immune system dysfunction, which improved the density of the network in our subsequent network meta-analysis (NMA). In addition to network meta-analysis, we also conducted a pairwise meta-analysis of the results of each random effect before constructing the network to add validity to certain findings. The outcomes of interest not only included common clinical manifestations such as mortality, hospitalization, ICU admission, MV requirement and adverse reactions in previous meta-analyses [150-153], but also different specific symptom assessments such as improvement of olfactory score, venous thrombotic events (VTEs), arterial thrombotic events (ATEs) and major bleeding events (MBEs). We even carried out a separate small NMA to evaluate the efficacy and safety of different doses of anticoagulants for cardiovascular symptom associated with COVID-19 as a complementary demonstration of some of our findings [154]. Olfactory training (OT), saline nasal irrigation (SNI), nitrilotriacetic acid trisodium (NAT), tetra sodium pyrophosphate (TSPP), mometasone furoate nasal spray (MFNS), sodium gluconate, palmitoylethanolamide and luteolin combined with olfactory training (OT), mometasone furoate nasal spray (MFNS) plus olfactory training (OT) and omega-3 fatty acids plus olfactory training (OT) proved to be effective treatments for post-COVID olfactory dysfunction in recent randomized controlled trials.

Our study has several limitations. First, some of the results were derived from a single study or studies with high RoB. To account for such weakness in evidence, we assessed the certainty of evidence for each outcome using the GRADE framework as summarized in Supplementary Appendix 2. Second, for certain treatment agents, many articles have been published among which only one or few have been included in our analysis. This is because we prospectively collected studies that adhered to predefined inclusion criteria, and studies that did not adequately account for confounding or those prone to significant bias were filtered out. Third, we included studies of pharmacological interventions of COVID-19 symptoms in addition to long-COVID symptoms. In order to construct a large network meta-analysis, the studies we included was not only limited to long COVID, but also involved a large number of trials on symptoms related to the COVID-19. Therefore, the results of our evaluation of potential treatments for the sequelae of COVID-19 are only of indirect referential value and still require to be confirmed by further clinical trials. Furthermore, we attempted to minimize biases by exclusively including observational studies that accounted for potential confounders and further conducted sensitivity analyses in which the same analysis was performed using only RCTs. Lastly, some of the results derived from this NMA lacks the support of pairwise meta-analysis. However, the methodological power of NMA is credible as empirical evidence supported that NMAs were 20% more likely to provide stronger evidence against the null hypothesis than conventional pairwise meta-analyses [155]. Accordingly, our NMA can offer meaningful implications for guiding management of COVID-19 and long COVID until future studies build up stronger evidence.

The gradient of evidence levels analyzed in this review may assist the decision-making of clinicians and policymakers. Although a number of studies reported consistent results on beneficial effect of olfactory therapy associated with long COVID, the evidence of treatments for long-COVID-related cardiovascular symptoms, organ failure, mental disorders and inflammatory response are either low or very low, due to the lack of direct and reliable randomized controlled trial evaluations. RCTs on these potential pharmacological interventions for long-COVID are required to confirm these findings and increase the level of evidence. For post-COVID-19 anosmia, saline nasal irrigation (SNI), nitrilotriacetic acid trisodium (NAT), tetra sodium pyrophosphate (TSPP), sodium gluconate, mometasone furoate nasal spray (MFNS) and olfactory training (OT) may effectively improve smell scores of long COVID as shown in our NMA from RCTs; there is evidence of associations for other agents from other clinical indicators or observational data that tocilizumab, dexamethasone, ivermectin nanosuspension mucoadhesive nasal spray, intranasal betamethasone sodium phosphate drops (IBSPD) and zinc sulfate may also provide clinical benefits of olfactory outcomes. In our another small NMA, therapeutic doses of anticoagulants proved more effective than prophylactic doses in preventing clots, but carried a greater risk of major bleeding in terms of safety. Among 36 published clinical trials directly involved in treatments for long COVID included in our evaluation, rivaroxaban proved to reduce both ATEs and VTEs for post-discharge thromboprophylaxis; Shugan Jieyu capsule (SGJYC), Shumian capsules (SMC), cognitive behavioural therapy (CBT) and neurofeedback therapy (NFBT) proved associated with improved clinical psychological outcomes of long COVID; hyperbaric oxygen therapy (HBOT) proved to induce neuroplasticity and improve neurocognitive functions of patients suffering from post-COVID condition; H2 inhalation might be a safe and effective way to accelerate early functional recovery in post-COVID-19 patients; leronlimab has been suggested to improve inflammatory response in long COVID. while posing both cardiac and noncardiac safety risks and warrant appropriate patient selection; azithromycin is related to dealing with hair loss after COVID-19. Only 12.8% of current evidence on pharmacological management of long COVID is on moderate/high evidence certainty and can therefore be confidently incorporated into practice and policy.

## DECLARATIONS

### Acknowledgments

None.

### Availability of data or materials

The materials of datasets used for the current study were accessible in the specific public database (as we describe in the Methods section). The analysis codes are available upon reasonable request from the corresponding author.

### Author Contributions

H.Z. had the idea for the article, designed the analysis, and prepared the manuscript. H.Z. performed the literature search and data analysis. F.J. and Z.J. supervised the studies, provided unique insights based on lived experience and specialist knowledge, respectively. H.Z. and F.J. revised the manuscript. Z.J. is the guarantor and accepts full responsibility for the work and/or the conduct of the study, had access to the data, and controlled the decision to publish. All authors reviewed and approved the manuscript.

### Funding

This research was funded by the Hunan Provincial Natural Science Foundation of China (No. 2023JJ30396).

### Ethics declarations

Not applicable.

## Conflicts of Interest

The authors declare no conflict of interest. The funders had no role in the design of the study; in the collection, analyses, or interpretation of data; in the writing of the manuscript; or in the decision to publish the results.

**Figure legends**

**Figure 1. Study selection.**

**Figure 2. Pairwise meta-analysis of pharmacological interventions compared with control (placebo/SOC) for all-cause mortality (45 studies) (A), hospital admission (10 studies) (B), hospital days (10 studies) (C), ICU admission (9 studies) (D), ICU days (14 studies) (E), MV requirement (12 studies) (F) and MV days (6 studies) (G).**

**Figure 3. Network of eligible comparisons for primary efficacy and safety outcomes.** The figure plots the network of eligible direct comparisons for all-cause mortality (41 studies) (A), hospital admission (8 studies) (B), hospital days (10 studies) (C), ICU admission (8 studies) (D), ICU days (14 studies) (E), MV requirement (11 studies) (F), MV days (6 studies) (G), adverse events (AE) (19 studies) (H) and serious adverse events (SAE) (10 studies) (I). The width of the lines is proportional to the number of trials comparing every pair of treatments, and the size of every node is proportional to the number of randomized participants.

**Figure 4. Network meta-analysis of all-cause mortality (blue), adverse events (light red) serious adverse events (red).** Pharmacological treatments are reported in alphabetical order. Comparisons should be read from left to right. All-cause mortality and safety estimates are located at the intersection between the column-defining and the row-defining treatment. For all-cause mortality, ORs above 1 favor the column-defining treatment. For safety, ORs above 1 favor the row-defining treatment. We incorporated the GRADE judgments in the figure. Estimates in gray have a very low or low certainty of evidence.

**Figure 5. Network of eligible comparisons for improvement of smell scores (13 studies).**

**Figure 6. Network of eligible comparisons for efficacy and safety outcomes of anticoagulant drugs.** The figure plots the network of eligible direct comparisons for arterial thrombotic events (ATEs) (5 studies) (A), venous thrombotic events (VTEs) (5 studies) (B) and major bleeding events (MBEs) (6 studies) (C). The width of the lines is proportional to the number of trials comparing every pair of treatments, and the size of every node is proportional to the number of randomized participants.

**Figure 7. Forest plot of eligible comparisons for efficacy and safety outcomes of different doses of anticoagulants.** The figure plots the network of eligible direct comparisons for arterial thrombotic events (ATEs) (7 studies) (A), venous thrombotic events (VTEs) (8 studies) (B) and major bleeding events (MBEs) (6 studies) (C). The width of the lines is proportional to the number of trials comparing every pair of treatments, and the size of every node is proportional to the number of randomized participants.

**TABLE 1 | Characteristics of included randomized controlled trials.**

**Supplementary information: Supplementary Appendix 1-5**

**Figure 1.**

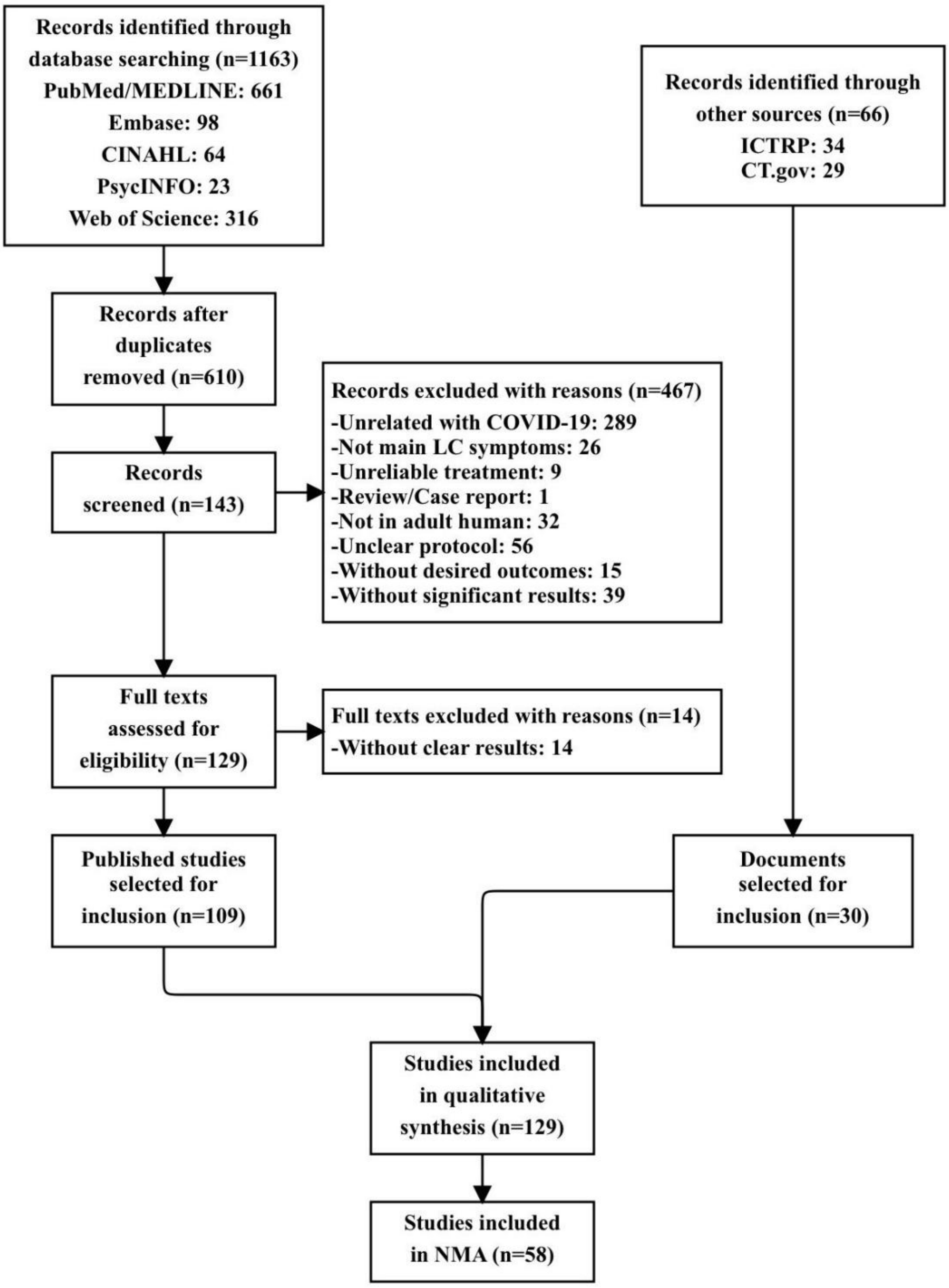
Records identified through database searching (n=1163)
PubMed/MEDLINE: 661
Embase: 98
CINAHL: 64
PsycINFO: 23
Web of Science: 316
Records identified through other sources (n=66)
ICTRP: 34
CT.gov: 29
Records after duplicates removed (n=610)
Records excluded with reasons (n=467)
-Unrelated with COVID-19: 289
-Not main LC symptoms: 26
-Unreliable treatment: 9
-Review/Case report: 1
-Not in adult human: 32
-Unclear protocol: 56
-Without desired outcomes: 15
-Without significant results: 39
Records screened (n=143)
Full texts assessed for eligibility (n=129)
Full texts excluded with reasons (n=14)
-Without clear results: 14
Published studies selected for inclusion (n=109)
Documents selected for inclusion (n=30)
Studies included in qualitative synthesis (n=129)
Studies included in NMA (n=58)

**Figure 2.**

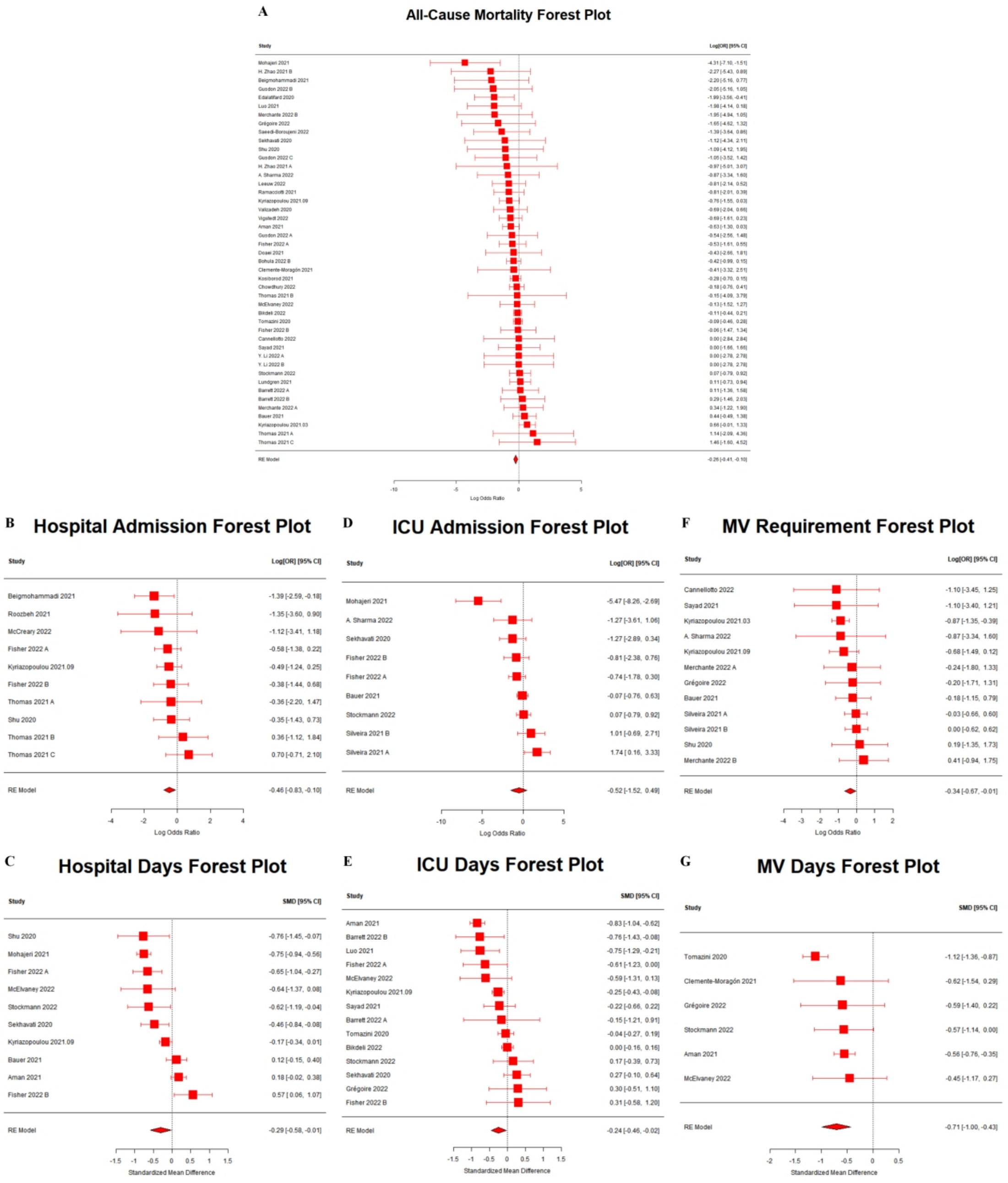
A
All-Cause Mortality Forest Plot
B
Hospital Admission Forest Plot
C
Hospital Days Forest Plot
D
ICU Admission Forest Plot
E
ICU Days Forest Plot
F
MV Requirement Forest Plot
G
MV Days Forest Plot

**Figure 3.**

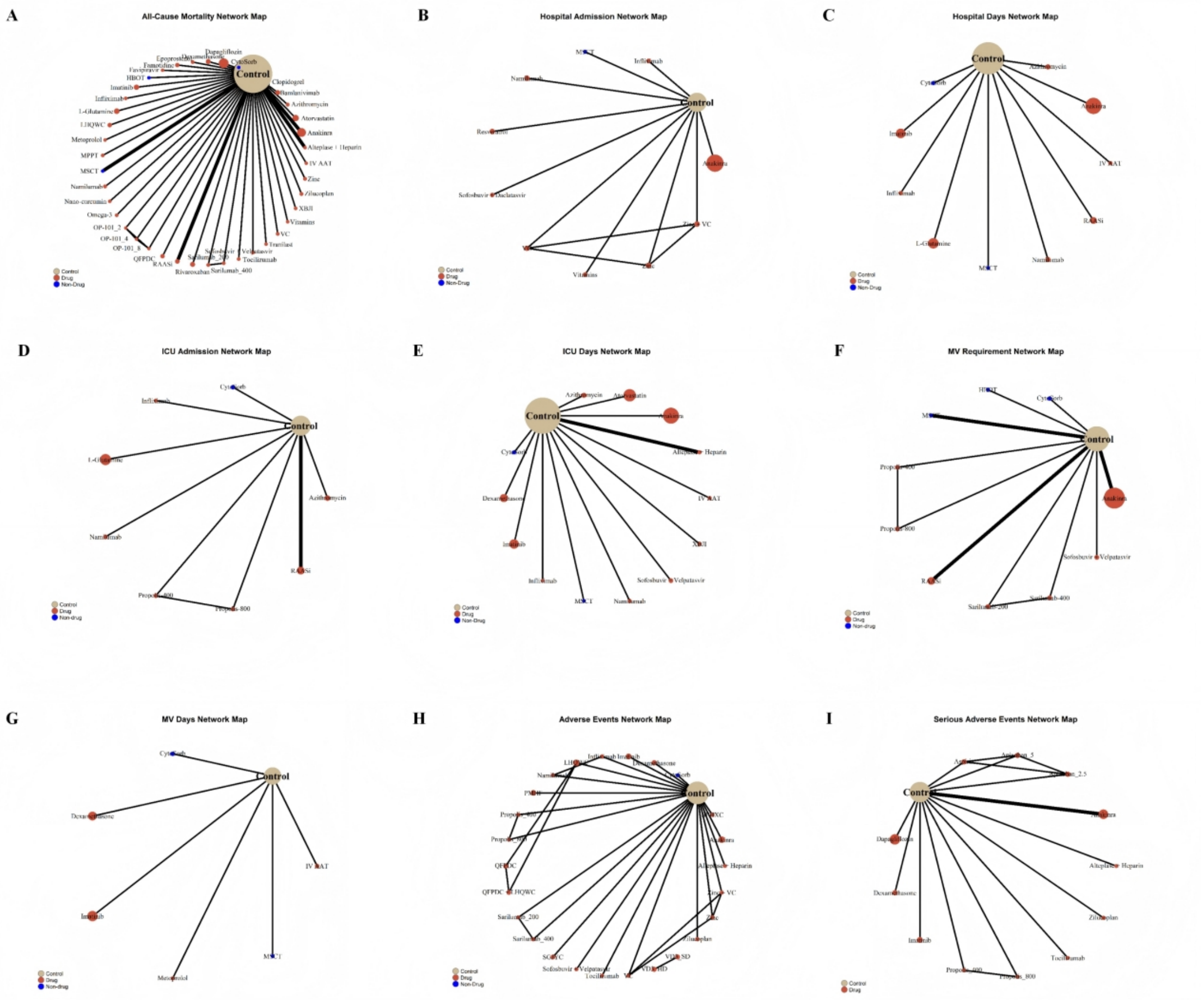
A
All-Cause Mortality Network Map
Control
B
Hospital Admission Network Map
Control
C
Hospital Days Network Map
Control
D
ICU Admission Network Map
Control
E
ICU Days Network Map
Control
F
MV Requirement Network Map
Control
G
MV Days Network Map
Control
H
Adverse Events Network Map
Control
I
Serious Adverse Events Network Map
Control

# Figure 4.

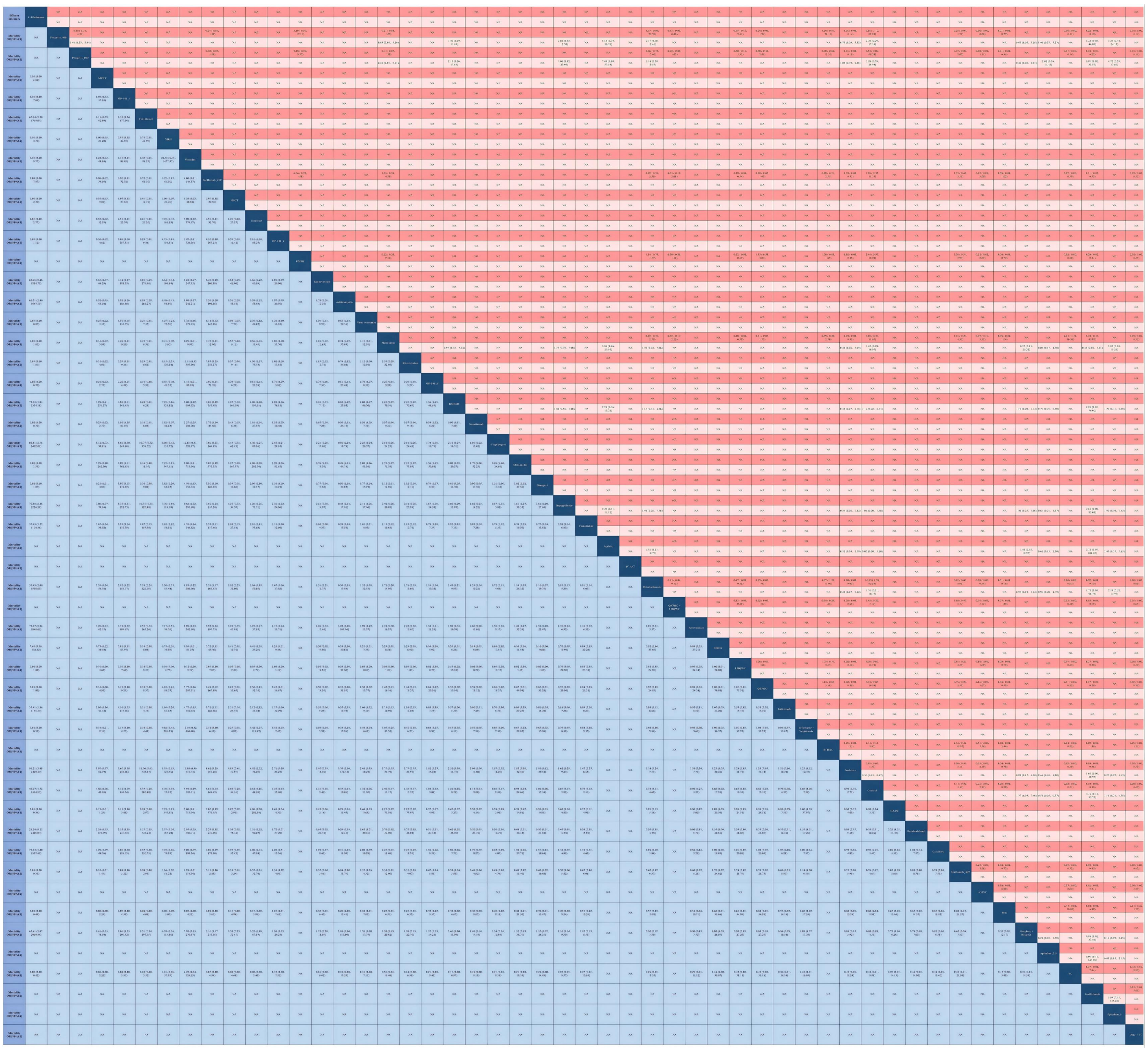

**Figure 5.**

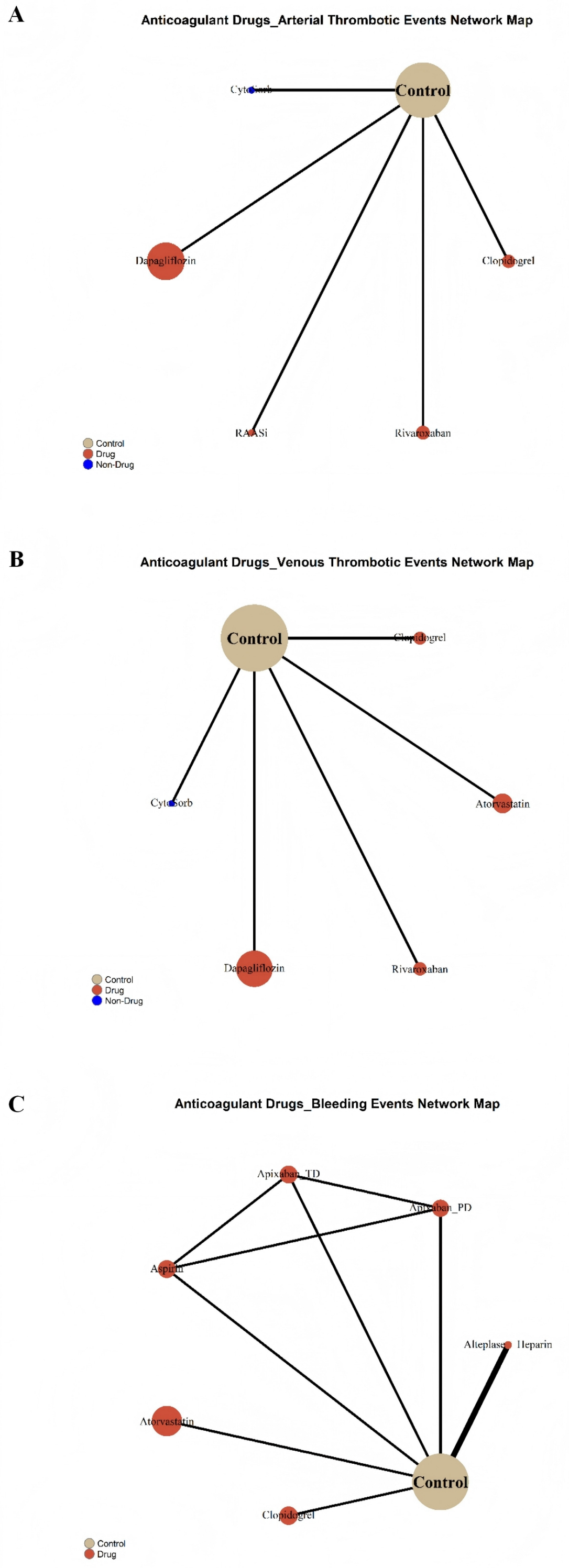

Figure 6.

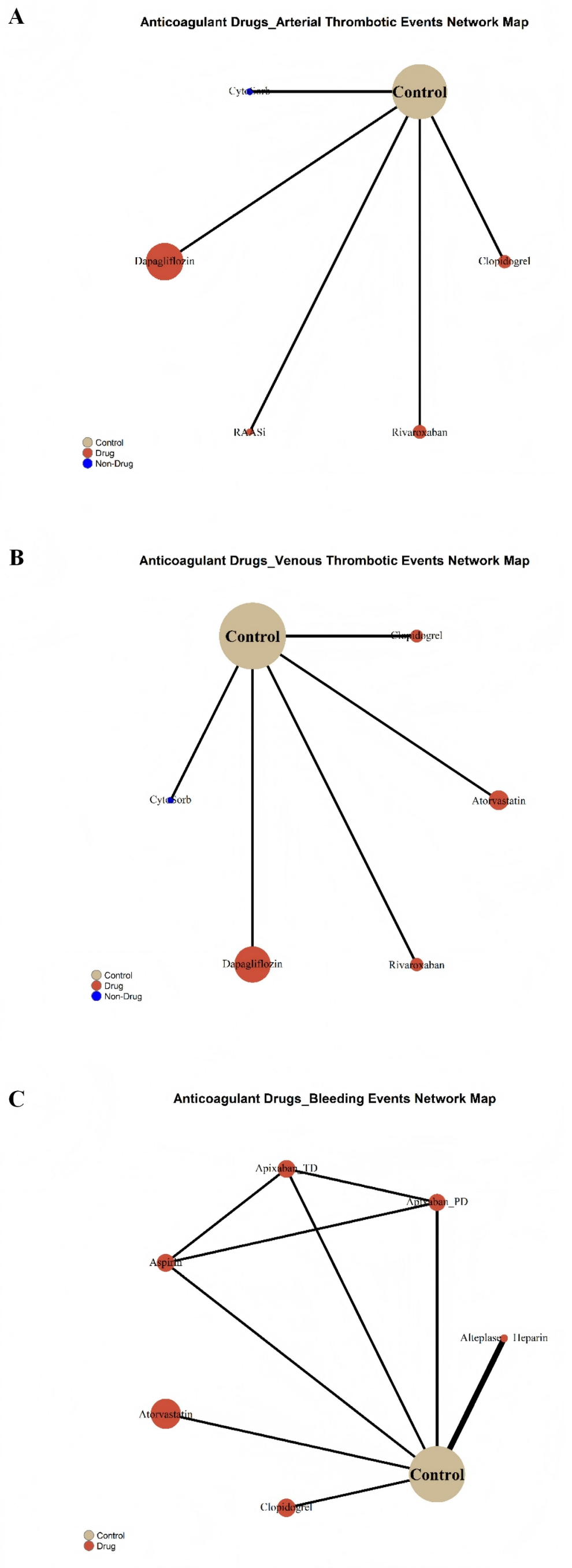

**Figure 7.**

**A**

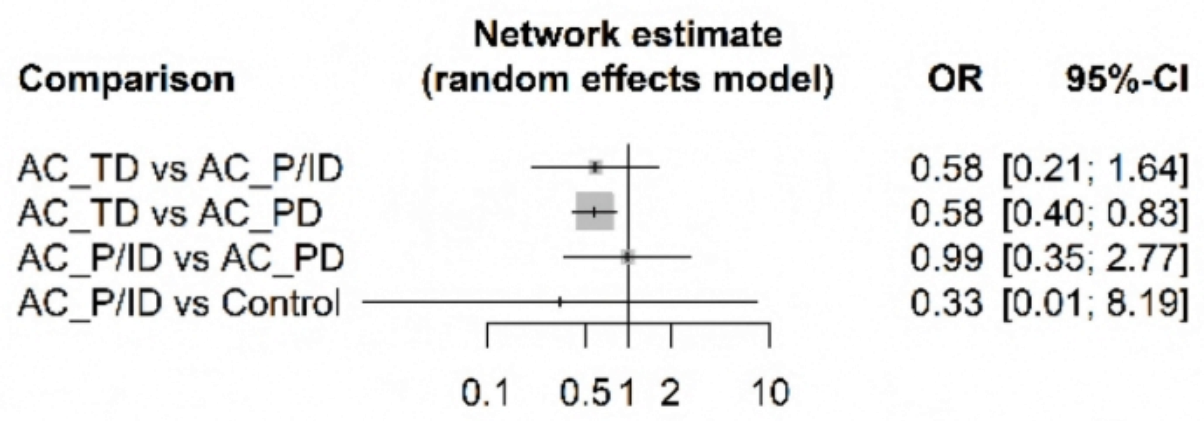

Comparison
Network estimate (random effects model)
OR
95%-CI
AC_TD vs AC_P/ID 0.58 [0.21; 1.64]
AC_TD vs AC_PD 0.58 [0.40; 0.83]
AC_P/ID vs AC_PD 0.99 [0.35; 2.77]
AC_P/ID vs Control 0.33 [0.01; 8.19]
0.1 0.5 1 2 10

**B**

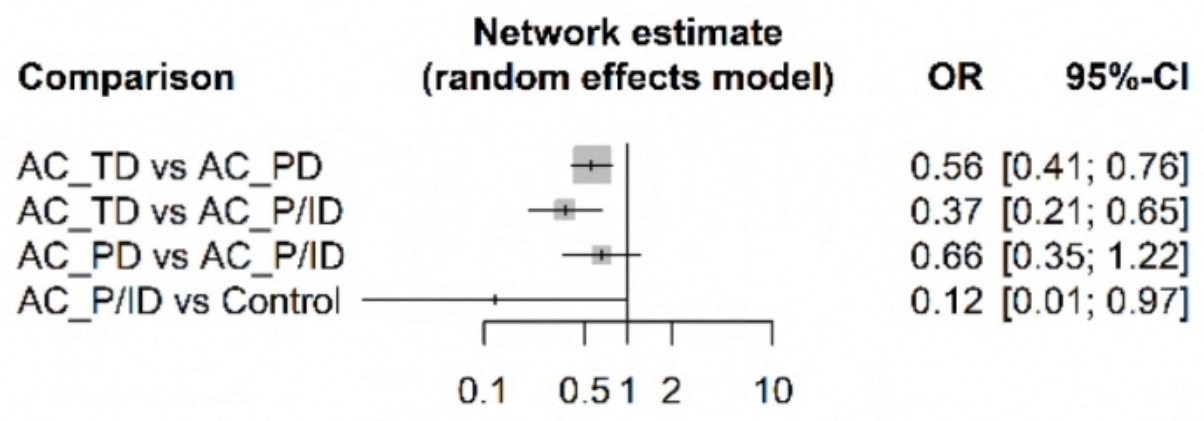

Comparison
Network estimate (random effects model)
OR
95%-CI
AC_TD vs AC_PD 0.56 [0.41; 0.76]
AC_TD vs AC_P/ID 0.37 [0.21; 0.65]
AC_PD vs AC_P/ID 0.66 [0.35; 1.22]
AC_P/ID vs Control 0.12 [0.01; 0.97]
0.1 0.5 1 2 10

**C**

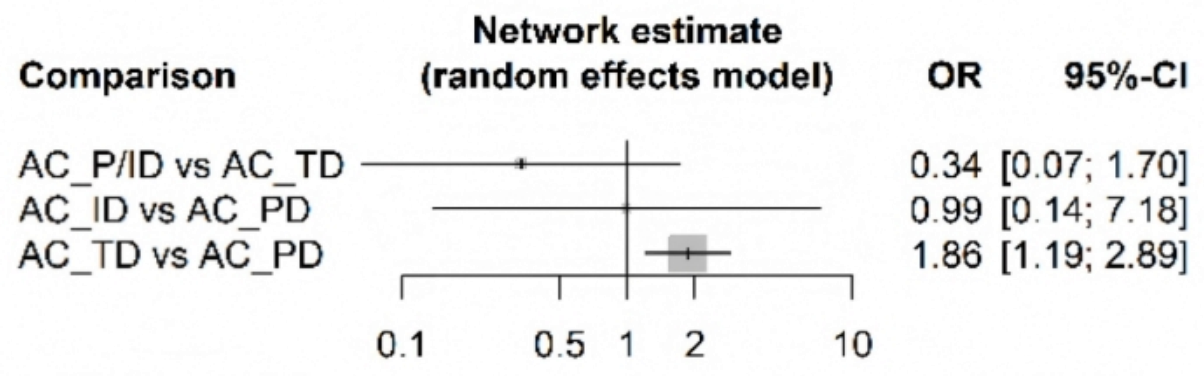

Comparison
Network estimate (random effects model)
OR
95%-CI
AC_P/ID vs AC_TD 0.34 [0.07; 1.70]
AC_ID vs AC_PD 0.99 [0.14; 7.18]
AC_TD vs AC_PD 1.86 [1.19; 2.89]
0.1 0.5 1 2 10

## Table 1.

| No. | Year | Author | Study design | Intervention type | Sample size | Intervention/Case group | Intervention abbreviation | Control group | Control abbreviation | Symptom cluster1 | Symptom cluster2 | Follow-up (days) | Diagnosis of COVID-19 | Post-COVID-19 (Y/N) | Disease severity (Severe/Moderate/ Mild) | Hospitalized (Y/N) | ICU (Y/N) | Comment |
|---|---|---|---|---|---|---|---|---|---|---|---|---|---|---|---|---|---|---|
| 1 | 2022 | A. Sharma | RCT, open-label | Pharmacological intervention | 46 (21:25) | discontinue RAASi for 7 days | RAASi | continue RAASi | SOC | Immune dysregulation and cardiovascular symptoms | Systemic inflammation and thrombosis | 7 | RT-PCR test | N | mild/moderate | Y | N | / |
| 2 | 2021 | Abdelalim | RCT, open-label | Pharmacological intervention, intra-nasal spray | 100 (50:50) | receive topical corticosteroid (mometasone furoate) nasal spray in an appropriate dose of 2 puff (100 μg) once daily in each nostril with olfactory training for 3 weeks | MFNS | receive olfactory training (OT) | OT | Smell/taste disorder | Olfactory dysfunction | 21 | RT-PCR test | Y | N | / | N | / |
| 3 | 2022.05 | Abdelazim | RCT, double-blind | Pharmacological intervention, intra-nasal spray | 58 (29:29) | receive a single dose of 0.1 mL of intranasal 2% NAT in phosphate buffer with a pH of 6 three times daily for 1 month | NAT | receive placebo | Placebo | Smell/taste disorder | Olfactory dysfunction | 30 | RT-PCR test | N | / | / | N | / |
| 4 | 2022.08 | Abdelazim | RCT, double-blind | Pharmacological intervention, intra-nasal spray | 64 (32:32) | receive intranasal spray 1% TSPP in borate buffer solution with a pH of 8 for a month | TSPP | receive placebo | Placebo | Smell/taste disorder | Olfactory dysfunction | 30 | RT-PCR test | Y | N | / | N | / |
| 5 | 2021 | Abdelmaksoud | a longitude observational study | Pharmacological intervention | 105 (49:56) | receive zinc therapy (220 mg zinc sulfate equivocal to 50 mg elemental zinc twice daily) plus the Egyptian protocol of treatment of COVID-19 | Zinc | receive standard-of-care (SOC) treatment | SOC | Smell/taste disorder | Olfactory and gustatory dysfunction | 22 | rRT-PCR test | N | / | / | N | / |
| 6 | 2021 | Altay | RCT, open-label, phase 2 | Pharmacological intervention | 93 (71:22) | receive combined metabolic activators (CMAs) with standard of care | CMAs | receive placebo | Placebo | Immune dysregulation | Inflammatory response | 14 | RT-PCR test | N | mild/moderate | N | N | / |
| 7 | 2021 | Altay | RCT, double-blind, phase 3 | Pharmacological intervention | 304 (229:75) | receive combined metabolic activators (CMAs) with standard of care | CMAs | receive placebo | Placebo | Immune dysregulation | Inflammatory response | 14 | RT-PCR test | N | mild/moderate | N | N | / |
| 8 | 2021 | Aman | RCT, double-blind | Pharmacological intervention | 385 (197:188) | receive a dose of 800 mg imatinib on day 0, followed by 400 mg once daily on days 1–9 | Imatinib | receive placebo | Placebo | Organ damage/failure | Acute respiratory distress syndrome (ARDS) and hypoxaemic respiratory failure | 28 | RT-PCR test | N | severe | Y | N | / |
| 9 | 2022 | An | RCT, double-blind | Pharmacological intervention, traditional Chinese medicine (TCM) treatment | 196 (99:97) | receive Shugan Jieyu capsule | SGJYC | receive placebo | Placebo | Mental disorders | Depression, anxiety, insomnia and fatigue | 42 | Unclear | Y | N | N | N | / |
| 10 | 2022 | Annweiler | RCT, open-label | Pharmacological intervention, nutritional supplement | 254 (127:127) | receive high-dose (400,000 IU) vitamin D3 supplementation | VD3_HD | receive standard-dose (50,000 IU) vitamin D3 supplementation | VD3_SD | Immune dysregulation | Inflammatory response | 28 | RT-PCR test or chest CT scan | N | / | Y | N | at-risk older patients |
| 11 | 2021 | Aref | RCT, double-blind | Pharmacological intervention, intra-nasal spray | 114 (57:57) | receive ivermectin nanosuspension nasal spray twice daily plus the Egyptian protocol of treatment for COVID-19 | Ivermectin | receive standard-of-care (SOC) treatment | SOC | Smell/taste disorder | Olfactory dysfunction | 7 | RT-PCR test | N | mild | N | N | / |
| 12 | 2022 | Aref | RCT, open-label | Pharmacological intervention, intra-nasal spray | 96 (49:47) | receive ivermectin nanosuspension mucoadhesive nasal spray (two puffs per day) | Ivermectin | receive placebo | Placebo | Smell/taste disorder | Olfactory and gustatory dysfunction | 90 | RT-PCR test | Y | N | N | N | / |
| 13 | 2022 | Arianna | RCT, double-blind | Pharmacological intervention, nutritional supplement | 185 (130:55) | receive daily oral supplementation with ultramicronized PEA-LUT 770 mg plus olfactory training | Palmidrol + Luteolin + OT | receive olfactory training (OT) | OT | Smell/taste disorder | Olfactory dysfunction | 90 | RT-PCR test | Y | N | N | N | / |
| 14 | 2022 | Barrett | RCT, open-label, phase 2 | Pharmacological intervention | 14 (6:8) | receive a tPA drip (50-mg tPA IV bolus, followed by tPA drip 2 mg/h plus heparin 500 units/h over 24 h, then heparin to maintain aPTT of 60-80 s for 7 days) | Alteplase + Heparin | receive standard-of-care (SOC) treatment | SOC | Organ damage/failure and cardiovascular symptoms | Respiratory failure and thrombosis | 7 | chest CT scan | N | / | Y | Y | / |
| 15 | 2022 | Barrett | RCT, open-label, phase 1 | Pharmacological intervention | 36 (19:17) | receive a tPA bolus (50-mg tPA IV bolus followed by 7 days of heparin; goal activated partial thromboplastin time [aPTT], 60-80 s) | Alteplase + Heparin | receive standard-of-care (SOC) treatment | SOC | Organ damage/failure and cardiovascular symptoms | Respiratory failure and thrombosis | 7 | chest CT scan | N | / | Y | Y | / |
| 16 | 2021 | Bauer | RCT, open-label | Pharmacological intervention | 204 (104:100) | discontinue RAS inhibitors for 30 days | RAASi | continue RAS inhibitors for 30 days | SOC | Organ damage/failure | Multi-organ failure | 30 | RT-PCR test | N | / | / | / | / |
| 17 | 2021 | Beigmohammadi | RCT, single-blind | Pharmacological intervention, nutritional supplement | 60 (30:30) | receive 25,000 IU daily of vitamins A, 600,000 IU once during the study of D, 300 IU twice daily of E, 500 mg four times daily of C, and one amp daily of B complex for 7 days | Vitamins | receive placebo | Placebo | Immune dysregulation | Inflammatory response | 7 | RT-PCR test | N | severe | Y | Y | / |
| 18 | 2022 | Bikdeli | RCT, double-blind | Pharmacological intervention | 587 (290:297) | receive atorvastatin 20 mg orally once daily for 30 days | Atorvastatin | receive placebo | Placebo | Cardiovascular symptoms | Thrombosis | 30 | RT-PCR test | N | / | Y | Y | / |
| 19 | 2022 | Bohula | RCT, open-label | Pharmacological intervention, anticoagulant therapy | 382 (191:191) | receive full-dose prophylactic anticoagulation (FDPAC) | AC_TD | receive standard-dose prophylactic anticoagulation (SDPAC) | AC_PD | Cardiovascular symptoms and immune dysregulation | Thrombosis and inflammatory response | 28 | Unclear | N | severe | Y | Y | / |
| 20 | 2022 | Bohula | RCT, open-label | Pharmacological intervention, anticoagulant therapy | 290 (150:140) | receive clopidogrel therapy | Clopidogrel | receive standard-of-care (SOC) treatment | SOC | Cardiovascular symptoms and immune dysregulation | Thrombus and inflammatory response | 28 | Unclear | N | severe | Y | Y | / |
| 21 | 2022 | Botek | RCT, single-blind | Non-pharmacological intervention | 54 (27:27) | receive 14 days of H2 inhalation | H2 | receive placebo | Placebo | Mental disorders | Physical and respiratory function | 14 | RT-PCR test | N | / | N | N | / |
| 22 | 2020 | Bronte | a longitude observational study | Pharmacological intervention | 76 (20:56) | receive 4 mg baricitinib twice daily for 2 days, followed by 4 mg per day for the remaining 7 days | Baricitinib | No | No | Immune dysregulation | Inflammatory response | 7 | RT-PCR test | N | severe | Y | / | / |
| 23 | 2022 | Cannellotto | RCT, open-label | Non-pharmacological intervention, hyperbaric oxygen therapy (HBOT) | 40 (20:20) | receive hyperbaric oxygen (HBO2) treatment (≥5 sessions of 90 min of hyperbaric oxygen therapy administered once daily using Revitalair technology (1.45 ATA) with an inspired fraction close to 100% of oxygen) | HBOT | receive standard-of-care (SOC) treatment | SOC | Organ damage/failure | Hypoxaemia | 30 | RT-PCR test | N | severe | Y | Y | / |

# Table 1.

| No. | Year | Author | Study design | Intervention type | Sample size | Intervention/Case group | Intervention abbreviation | Control group | Control abbreviation | Symptom cluster1 | Symptom cluster2 | Follow-up (days) | Diagnosis of COVID-19 | Post-COVID-19 (Y/N) | Disease severity (Severe/Moderate/ Mild) | Hospitalized (Y/N) | ICU (Y/N) | Comment |
|---|---|---|---|---|---|---|---|---|---|---|---|---|---|---|---|---|---|---|
| 24 | 2022 | Cash | a longitude observational study | Pharmacological intervention, nutritional supplement | 43 (22:21) | receive 500 mg anhydrous enol-oxaloacetate (AEO) capsules BID or 1000 mg AEO BID for 6 weeks | AEO BID | self (historical placebo) | Placebo | Mental disorders | Fatigue | 42 | Unclear | N | / | N | N | / |
| 25 | 2022 | Catalogna | RCT, double-blind | Non-pharmacological intervention, hyperbaric oxygen therapy (HBOT) | 73 (37:36) | receive 40 daily sessions of HBOT | HBOT | receive standard-of-care (SOC) treatment | SOC | Mental disorders | Cognitive impairment | 40 | RT-PCR test | Y | N | / | N | / |
| 26 | 2022 | Chen | RCT, double-blind | Pharmacological intervention, traditional Chinese medicine (TCM) treatment | 129 (64:65) | receive Bufei Huoxue (BFHX) orally three times a day (1.4 g/dose) for 90 days | BFHXC | receive placebo | Placebo | Organ damage/failure | Cardiopulmonary symptoms | 90 | chest CT scan | Y | N | Y | / | / |
| 27 | 2022 | Chowdhury | RCT, single-blind | Pharmacological intervention, H2 receptor blocker | 208 (104:104) | received famotidine (Famotac 20 mg oral tablet formulation) every 8 h, 30 min before the meal; 40 mg in the case of < 60 kg, and 60 mg in the case of > 60 kg body weight for 30 days; other treatments included remdesivir, tocilizumab, dexamethasone, a broad-spectrum antibiotic (meropenem), proton pump inhibitor, ascorbic acid, cholecalciferol, zinc, bronchodilators and oxygen support | Famotidine | receive standard-of-care (SOC) treatment | SOC | Overall | Overall | 30 | RT-PCR test | N | severe | Y | Y | / |
| 28 | 2021 | Clemente-Moragón | RCT, open-label, a pilot trial | Pharmacological intervention | 20 (12:8) | receive metoprolol (15 mg daily for 3 days) | Metoprolol | receive standard-of-care (SOC) treatment | SOC | Organ damage/failure | Acute respiratory distress syndrome (ARDS) and respiratory failure | 3 | RT-PCR test | N | severe | Y | Y | / |
| 29 | 2021 | Connors | RCT, double-blind | Pharmacological intervention | 657 (164:165:164:164) | receive aspirin (81 mg orally once daily), prophylactic-dose apixaban (2.5 mg orally twice daily) or therapeutic-dose apixaban (5 mg orally twice daily) for 45 days | Aspirin/Apixaban (AC_PD/AC_TD) | receive placebo | Placebo | Cardiovascular symptoms | Thrombosis | 45 | Unclear | N | / | N | N | / |
| 30 | 2022 | Corral | RCT, double-blind | Non-pharmacological intervention, exercise training | 44 (22:22) | receive inspiratory muscle training (IMT) for 4 weeks | IMT | receive standard-of-care (SOC) treatment | SOC | Mental disorders | Fatigue | 56 | RT-PCR test | Y | N | N | N | / |
| 31 | 2022 | Corral | RCT, double-blind | Non-pharmacological intervention, exercise training | 44 (22:22) | receive respiratory (inspiratory/expiratory) muscle training (RMT) for 4 weeks | RMT | receive standard-of-care (SOC) treatment | SOC | Mental disorders | Fatigue | 56 | RT-PCR test | Y | N | N | N | / |
| 32 | 2021 | D'Ascanio | RCT, single-blind, a pilot trial | Pharmacological intervention, nutritional supplement | 12 (7:5) | receive olfactory training/stimulation through Sniffin' Sticks, administered as in control plus daily treatment with PEA/Luteolin oral supplement | Palmidrol + Luteolin + OT | receive olfactory training (OT) | OT | Smell/taste disorder | Olfactory dysfunction | 30 | RT-PCR test | Y | N | N | N | / |
| 33 | 2021 | Della-Torre | a non-randomized controlled observational study | Pharmacological intervention | 140 (52:88) | receive biologic agents in addition to local standard of care with anakinra | Anakinra | receive standard-of-care (SOC) treatment | SOC | Organ damage/failure and immune dysregulation | Severe pneumonia and hyper-inflammation | 186 | RT-PCR test | N | severe | Y | N | / |
| 34 | 2021 | Della-Torre | a non-randomized controlled observational study | Pharmacological intervention | 133 (30:103) | receive biologic agents in addition to local standard of care with sarilumab | Sarilumab | receive standard-of-care (SOC) treatment | SOC | Organ damage/failure and immune dysregulation | Severe pneumonia and hyper-inflammation | 186 | RT-PCR test | N | severe | Y | N | / |
| 35 | 2021 | Della-Torre | a non-randomized controlled observational study | Pharmacological intervention | 93 (25:68) | receive biologic agents in addition to local standard of care with tocilizumab | Tocilizumab | receive standard-of-care (SOC) treatment | SOC | Organ damage/failure and immune dysregulation | Severe pneumonia and hyper-inflammation | 186 | RT-PCR test | N | severe | Y | N | / |
| 36 | 2021 | Doaei | RCT, double-blind | Pharmacological intervention, nutritional supplement | 101 (28:73) | received one capsule of 1000 mg omega-3 daily containing 400 mg EPAs and 200 mg DHAs for 14 days through adding the sup- plement to their enteral formula | Omega-3 fatty acids | receive standard-of-care (SOC) treatment | SOC | Immune dysregulation | Inflammatory response | 30 | RT-PCR test | N | severe | Y | Y | / |
| 37 | 2021 | Domênico | RCT, double-blind | Non-pharmacological intervention, auxiliary treatment | 106 (63:43) | receive 1% H2O2 for gargling, and 0.5% H2O2 for a nasal wash, in which patients gargled with a solution composed of 1% H2O2 and mint essence for 30 seconds, 3 times a day, for a 7-day period | H2O2 | receive placebo | Placebo | Overall | Overall | 7 | RT-PCR test | N | severe | Y | Y | / |
| 38 | 2020 | Edalatifard | RCT, single-blind | Pharmacological intervention, intravenous injection | 62 (34:28) | receive methylprednisolone pulse (intravenous injection, 250 mg/day for 3days) | MPPT | receive standard-of-care (SOC) treatment | SOC | Immune dysregulation and organ damage/failure | Inflammatory response and acute respiratory distress syndrome (ARDS) | 7 | RT-PCR test and chest CT scan | N | severe | / | / | / |
| 39 | 2021 | Fan | RCT, single-blind | Non-pharmacological intervention, cognitive behavioral therapy (CBT) | 111 (56:55) | receive NET and personalized psychological intervention | NET | receive personalized psychological intervention | SOC | Mental disorders | Post-traumatic stress symptoms (PTSS) and Post-traumatic stress disorder (PTSD) | 180 | Unclear | Y | N | N | N | / |
| 40 | 2022 | Farnoosh | RCT, double-blind | Pharmacological intervention | 44 (24:20) | receive standard of care plus melatonin at a dose of 3 mg three times daily for 14 day | Melatonin | receive standard-of-care (SOC) treatment | SOC | Immune dysregulation | Inflammatory response | 28 | RT-PCR and chest X-ray or CT scan | N | mild/moderate | Y | N | / |
| 41 | 2022 | Feng | RCT, double-blind | Pharmacological intervention, traditional Chinese medicine (TCM) treatment | 194 (97:97) | receive YDJDG (composed of Ephedrae Herba, Gypsum Fibrosum, Mori Cortex, Scutellariae Radix, Lepidii Semen, Lonicerae Japonicae Flos, Scrophulariae Radix, Moutan Cortex, Rehmanniae Radix, Atractylodis Macrocephalae Rhizoma, and Cimicifugae Rhizoma) therapy | YDJDG | receive standard-of-care (SOC) treatment | SOC | Immune dysregulation | Inflammatory response | 14 | Unclear | N | / | N | N | / |
| 42 | 2022 | Fessler | RCT, double-blind | Pharmacological intervention, nutritional supplement | 60 (30:30) | receive the active ingredient (ingest 600 mg Levagen+ twice daily (LEV)) for 4 weeks | Palmidrol | receive placebo | Placebo | Immune dysregulation | Inflammatory response | 28 | RT-PCR or antigen test | N | / | N | N | unvaccinated adults |
| 43 | 2022 | Fisher | RCT, open-label, phase 2 | Pharmacological intervention, intravenous injection | 57 (28:29) | receive a single intravenous dose of 5 mg/kg given over 2 h on day 1 | Infliximab | receive standard-of-care (SOC) treatment | SOC | Immune dysregulation | Inflammatory response | 28 | chest x-ray or CT scan, with or without RT-PCR test | N | / | Y | / | / |
| 44 | 2022 | Fisher | RCT, open-label, phase 2 | Pharmacological intervention, intravenous injection | 97 (52:45) | receive a single intravenous dose of 150 mg given over 1 h on day 1 | Namilumab | receive standard-of-care (SOC) treatment | SOC | Immune dysregulation | Inflammatory response | 28 | chest x-ray or CT scan, with or without RT-PCR test | N | / | Y | / | / |

# Table 1.

| No. | Year | Author | Study design | Intervention type | Sample size | Intervention/Case group | Intervention abbreviation | Control group | Control abbreviation | Symptom cluster1 | Symptom cluster2 | Follow-up (days) | Diagnosis of COVID-19 | Post-COVID-19 (Y/N) | Disease severity (Severe/Moderate/ Mild) | Hospitalized (Y/N) | ICU (Y/N) | Comment |
|---|---|---|---|---|---|---|---|---|---|---|---|---|---|---|---|---|---|---|
| 45 | 2022 | Gaylis | RCT, open-label | Pharmacological intervention, monoclonal antibody, CCR5-binding humanized immunoglobulin G4 monoclonal antibody | 65 (28:27) | receive weekly subcutaneous doses of either leronlimab (700 mg) for 8 weeks | Leronlimab | receive placebo | Placebo | Immune dysregulation | Inflammatory response | 56 | Unclear | Y | N | / | N | / |
| 46 | 2021 | Goligher | RCT, open-label | Pharmacological intervention, anticoagulant therapy | 1003 (536:567) | receive therapeutic-dose anticoagulation with unfractionated or low-molecular-weight heparin | LMWH (AC_TD) | receive usual-care pharmacologic thromboprophylaxis | AC_PD | Cardiovascular symptoms | Thrombosis | 21 | RT-PCR test | N | severe | Y | / | / |
| 47 | 2022 | Grégoire | RCT, open-label, phase 1/2 | Pharmacological intervention, mesenchymal stem cell (MSC) therapy | 32 (8:24) | receive 3 infusions of 1.5-3 × 106 BM-MSCs/kg, at 3 (± 1) day intervals | MSCT | receive standard-of-care (SOC) treatment | SOC | Organ damage/failure | Acute respiratory distress syndrome (ARDS) | 60 | chest CT scan and RT-PCR test | N | severe | Y | / | / |
| 48 | 2022 | Gupta | RCT, triple-label, phase 2 | Pharmacological intervention, intra-nasal spray | 51 (26:25) | receive 84 capsules of theophylline (400 mg/capsule) | SNI + theophylline | receive placebo | Placebo | Smell/taste disorder | Olfactory dysfunction | 42 | Unclear | Y | N | N | N | / |
| 49 | 2022 | Gusdon | RCT, double-blind, phase 2a | Pharmacological intervention, dendrimer nanotherapy | 24 (5:6:6:7) | receive OP-101 8 mg/kg or OP-101 4 mg/kg or OP-101 2 mg/kg | OP-101 | receive placebo | Placebo | Immune dysregulation | Inflammatory response | 30 | Unclear | N | severe | Y | / | / |
| 50 | 2021 | H. Zhao | RCT, open-label | Pharmacological intervention | 26 (14:7:5) | receive favipiravir + tocilizumab combination therapy | Favipiravir + Tocilizumab | receive favipiravir or tocilizumab monotherapy | Favipiravir/Tocilizumab | Immune dysregulation | Inflammatory response | 14 | chest CT scan | N | moderate/severe | Y | / | / |
| 51 | 2021 | Hashemian | Case series | Pharmacological intervention, mesenchymal stem cell (MSC) therapy | 11 | receive a total dose of 600 × 106 allogeneic human MSCs by intravenous infusions that were divided into three doses administered every other day | MSCT | No | No | Organ damage/failure | Acute respiratory distress syndrome (ARDS) | 60 | RT-PCR test or chest X-ray scan | N | severe | Y | Y | / |
| 52 | 2021 | Hassaniazad | RCT, triple-blind | Pharmacological intervention | 40 (20:20) | receive Sinacurcumin® soft gel 40 mg four times per day | Nano-curcumin | receive placebo | Placebo | Immune dysregulation | Inflammatory response | 7 | chest CT scan | N | / | Y | N | / |
| 53 | 2022 | Hawkins | RCT, double-blind | Pharmacological intervention, nutritional supplement | 40 (20:20) females | receive one 15 ml bottle containing a blend of essential oils extracted from the following plants: thyme (Thymus vulgaris), orange peel (Citrus sinensis), clove bud (Eugenia caryophyllus), and frankincense (Boswellia carterii) | Aromatherapy | receive placebo | Placebo | Mental disorders | Fatigue and insomnia | 14 | Unclear | Y | N | N | N | / |
| 54 | 2022 | Hernandez | RCT, open-label, a pilot trial | Pharmacological intervention, nutritional supplement | 58 (29:29) | receive olfactory training with omega-3 supplementation | Omega-3 + OT | receive olfactory training (OT) | OT | Smell/taste disorder | Olfactory dysfunction | 84 | Unclear | Y | N | N | N | / |
| 55 | 2021 | Hernandez-Cedeño | CCT, open-label | Pharmacological intervention, altered peptide ligand (APL) | 24 (11:13) | Seriously ill patients treated with oxygen therapy: receive 1 mg of CIGB-258 every 12 h; critically ill patients under invasive mechanical ventilation: receive the intravenous administration of 1 or 2 mg of CIGB-258 every 12 h. | CIGB-258 | No | No | Immune dysregulation | Hyperinflammatio n | 4 | RT-PCR test | N | severe | Y | / | / |
| 56 | 2022 | Hosseinpoor | RCT, double-blind | Pharmacological intervention, intra-nasal spray | 70 (35:35) | receive one puff of 0.05% wt/vol mometasone furoate intranasal spray on each side twice per day for 4 weeks | MFNS | receive placebo | Placebo | Smell/taste disorder | Olfactory dysfunction | 28 | RT-PCR test | Y | N | N | N | / |
| 57 | 2021 | Hu | RCT, open-label | Pharmacological intervention, traditional Chinese medicine (TCM) treatment | 284 (142:142) | receive usual treatment in combination with LH capsules (4 capsules, thrice daily) for 14 days | LHQWC | receive standard-of-care (SOC) treatment | SOC | Mental disorders | Fatigue | 14 | Unclear | N | / | Y | N | / |
| 58 | 2021 | J. Liu | RCT, open-label | Pharmacological intervention, traditional Chinese medicine (TCM) treatment | 149 (71:78) | receive 10 g (granules) twice daily for 14 days plus standard care | HSBDG | receive standard-of-care (SOC) treatment | SOC | Mental disorders | Fatigue | 14 | Unclear | N | / | Y | N | / |
| 59 | 2020 | J. Zhao | RCT, open-label | Pharmacological intervention, traditional Chinese medicine (TCM) treatment | 39 (15:24) | receive Integrated Chinese (the 5th edition recommendation's CM prescription extraorally for 2 weeks) and western medicine therapy | CMP | receive pure western medicine therapy | SOC | Organ damage/failur | Yidu-toxicity blocking lung syndrome | 29 | chest CT scan and rRT-PCR test | N | severe | / | N | / |
| 60 | 2022 | Jimeno-Almazán | RCT, open-label | Non-pharmacological intervention, exercise training | 39 (19:20) | receive supervised therapeutic training for 8 weeks (two supervised sessions per week comprised resistance training combined with aerobic training (moderate intensity variable training), plus a third day of monitored light intensity continuous training) | STE | receive standard-of-care (SOC) treatment | SOC | Mental disorders | Fatigue, anxiety and depression | 56 | RT-PCR or antigen test | Y | N | N | N | / |
| 61 | 2020.03 | K. Liu | RCT, open-label | Non-pharmacological intervention, muscle training | 51 (25:26) | use progressive muscle relaxation (PMR) technology for 30 min per day for 5 consecutive days | PMR | receive standard-of-care (SOC) treatment | SOC | Mental disorders | Anxiety and insomnia | 5 | Unclear | N | / | Y | N | / |
| 62 | 2020.04 | K. Liu | RCT, open-label | Non-pharmacological intervention, exercise training | 72 (36:36) | receive respiratory rehabilitation training (including: (1) respiratory muscle training; (2) cough exercise; (3) diaphragmatic training; (4) stretching exercise; and (5) home exercise) (2 sessions per week for 6 weeks) once a day for 10 min | RMT | receive standard-of-care (SOC) treatment | SOC | Mental disorders | Anxiety and depression | 42 | Unclear | N | / | Y | N | old patients |
| 63 | 2021 | Karimi | RCT, open-label | Pharmacological intervention, traditional Chinese medicine (TCM) treatment | 358 (184:174) | receive herbal remedies (polyherbal decoction every 8 hr and two herbal capsules every 12 hr) plus standard care for 7 days | PMHF | receive standard-of-care (SOC) treatment | SOC | Mental disorders | Fatigue | 7 | chest CT scan and RT-PCR test | N | / | Y | N | / |
| 64 | 2021 | Kasiri | RCT, double-blind | Pharmacological intervention, intra-nasal spray | 77 (39:38) | receive two puffs of mometasone furoate 0.05% nasal spray at an appropriate dose (100 A˚μg) twice daily in each nostril with olfactory training for four weeks | MFNS + OT | receive olfactory training (OT) | OT | Smell/taste disorder | Olfactory dysfunction | 28 | RT-PCR test or chest CT scan | N | / | N | N | / |
| 65 | 2021 | Khademi | RCT, open-label | Non-pharmacological intervention | 70 (35:35) | receive routine care and mandala colouring for 30 min/day in the evenings for six consecutive days | MC | receive standard-of-care (SOC) treatment | SOC | Mental disorders | Anxiety | 6 | Unclear | N | / | Y | N | / |
| 66 | 2021 | Kosiborod | RCT, double-blind, phase 3 | Pharmacological intervention, IL-1α/β inhibitor | 1250 (625:625) | receive dapagliflozin (10 mg daily orally) for 30 days | Dapagliflozin | receive placebo | Placebo | Organ damage/failure and cardiovascular symptoms | Multi-organ failure and thrombosis | 30 | Unclear | N | / | Y | N | / |

# Table 1.

| No. | Year | Author | Study design | Intervention type | Sample size | Intervention/Case group | Intervention abbreviation | Control group | Control abbreviation | Symptom cluster1 | Symptom cluster2 | Follow-up (days) | Diagnosis of COVID-19 | Post-COVID-19 (Y/N) | Disease severity (Severe/Moderate/ Mild) | Hospitalized (Y/N) | ICU (Y/N) | Comment |
|---|---|---|---|---|---|---|---|---|---|---|---|---|---|---|---|---|---|---|
| 67 | 2021.03 | Kyriazopoulou | RCT, open-label | Pharmacological intervention | 260 (130:130) | receive subcutaneous anakinra 100 mg once daily for 10 days in addition to standard of care | Anakinra | receive standard-of-care (SOC) treatment | SOC | Organ damage/failure | Severe respiratory failure (SRF) | 30 | Unclear | N | severe | Y | N | / |
| 68 | 2021.09 | Kyriazopoulou | RCT, double-blind, phase 3 | Pharmacological intervention | 594 (405:189) | receive anakinra | Anakinra | receive placebo | Placebo | Organ damage/failure | Respiratory failure | 28 | RT-PCR test | N | moderate/severe | Y | / | / |
| 69 | 2021 | L. Li | RCT, open-label | Pharmacological intervention, traditional Chinese medicine (TCM) treatment | 197 (98:99) | receive 3 (1.2 g) Shumian capsules at a time, 2 times a day for 2 weeks | SMC | receive placebo | Placebo | Mental disorders | Insomnia, anxiety and depression | 14 | RT-PCR test and chest CT scan | Y | N | N | N | / |
| 70 | 2021 | Lawler | RCT, open-label | Pharmacological intervention, anticoagulant therapy | 2231 (1181:1050) | receive therapeutic-dose anticoagulation with heparin | Heparin (AC_TD) | receive usual-care pharmacologic thromboprophylaxis | AC_PD | Cardiovascular symptoms and immune dysregulation | Thrombosis and inflammatory response | 21 | Unclear | N | moderate | Y | N | / |
| 71 | 2022 | Lechner | RCT, open-label | Non-pharmacological intervention, olfactory training | 63 (33:30) | receive a 12-week olfactory training (OT) | OT | receive standard-of-care (SOC) treatment | SOC | Smell/taste disorder | Olfactory dysfunction | 84 | RT-PCR test | Y | N | N | N | / |
| 72 | 2022 | Leeuw | RCT, open-label, phase 2 | Pharmacological intervention, C5 inhibitor | 78 (54:24) | receive standard of care with the C5 inhibitor zilucoplan daily for 14 days, under antibiotic prophylaxis | Zilucoplan | receive standard-of-care (SOC) treatment | SOC | Immune dysregulation and organ damage/failure | Systemic inflammation and hypoxemia | 14 | chest CT scan | N | / | / | / | / |
| 73 | 2022 | Londres | an observational study, phase 2/3 | Pharmacological intervention, monoclonal antibody | 41 (31 (severe) : 10 (moderate)) (7 (a single infusion) : 29 (two infusions) : 5 (three infusions)) | receive nimotuzumab together with the standard of care (SOC) (low-molecular-weight heparin, steroids and antibiotics) | Nimotuzumab | No | No | Immune dysregulation | Inflammatory response | 14 | RT-PCR test | N | moderate/severe | Y | / | / |
| 74 | 2022 | Luca | a longitude observational study | Pharmacological intervention, nutritional supplement | 69 (10:43:16) | previously exposed to olfactory training, receive PEA-LUT oral supplement and continued olfactory training | Palmidrol + Luteolin + OT | not previously exposed to olfactory training (Training-Naïve): Training-Naïve 1 group (PEA-LUT plus olfactory training): receive one sublingual sachet of PEA-LUT (700 mg + 70 mg) per day and olfactory training three times a day Training-Naïve 2 group (PEA-LUT alone): receive one sublingual sachet of PEA-LUT (700 mg + 70 mg) per day | OT/Palmidrol + Luteolin | Smell and mental disorders | Olfactory and memory dysfunction | 90 | RT-PCR test | Y | N | N | N | / |
| 75 | 2021 | Lundgren | RCT, double-blind | Pharmacological intervention, monoclonal antibody | 305 (157:148) | receive Bamlanivimab (7000mg) | Zilucoplan | receive placebo | Placebo | Overall | Overall | 90 | Unclear | N | moderate/severe | Y | / | / |
| 76 | 2021 | Luo | RCT, double-blind | Pharmacological intervention, traditional Chinese medicine (TCM) treatment, intravenous injection | 57 (29:28) | receive routine medication plus Xuebijing injection for 14 days, and 50 ml XBJ injection was diluted with 100 ml normal saline to 150 ml, every 12 h for 60 min | XBJI | receive placebo | Placebo | nmune dysregulatio | Inflammatory response | 14 | RT-PCR test | N | severe | Y | / | / |
| 77 | 2022.08 | M. H. and A. H. Abdelazim | RCT, double-blind | Pharmacological intervention, intra-nasal spray | 50 (25:25) | receive an intranasal spray of 1% sodium gluconate; 3 sprays for every nostril 3 times daily; a volume of 0.1 mL for every spray for a month | SG | receive placebo | Placebo | Smell/taste disorder | Olfactory dysfunction | 30 | RT-PCR test | Y | N | N | N | / |
| 78 | 2022 | Marashian | RCT, double-blind, a pilot trial | Non-pharmacological intervention | 52 (28:24) | receive Photobiomodulation (PBM) (620-635 nm light via 8 LEDs that provide an energy density of 45.40 J/cm2 and a power density of 0.12 W/cm2) twice daily for three days along with classical approved treatment | PBM | receive placebo | Placebo | nmune dysregulatio | Inflammatory response | 10 | RT-PCR test | N | mild/moderate | Y | / | / |
| 79 | 2021 | Marcos-Jubilar | RCT, open-label | Pharmacological intervention, anticoagulant therapy | 65 (32:33) | receive therapeutic-dose bemiparin (115 IU/kg daily) for 10 days | LMWH (Bemiparin) (AC_TD) | receive standard prophylaxis (bemiparin 3,500 IU daily) | LMWH (Bemiparin) AC_PD | lation and cardiovas | Inflammation response and thrombosis | 10 | RT-PCR test | N | mild/moderate | Y | N | / |
| 80 | 2022 | McCreary | RCT, open-label, phase 2 | Pharmacological intervention | 105 (52:53) | receive identically appearing bottles containing 60 identically appearing capsules of either>98% pure trans-resveratrol (from Japanese Knotweed Root, Polygonum cuspidatum extract) (500 mg per capsule) with vitamin D3 | Resveratrol + VD3 | receive vitamin D3 | VD3 | Overall | Overall | 60 | Unclear | N | mild | N | N | / |
| 81 | 2022 | McElvaney | RCT, double-blind, phase 2 (+ a pilot trial) | Pharmacological intervention | 36 (25:11) | receive a dose of 120 mg/kg IV AAT infusion, subsequently receive either weekly AAT or weekly placebo | IV AAT | receive placebo | Placebo | Organ damage/failure and immune dysregulation | Acute respiratory distress syndrome (ARDS) and inflammatory response | 28 | RT-PCR test | N | severe | Y | Y | / |
| 82 | 2022 | McNarry | RCT, open-label | Non-pharmacological intervention, exercise training | 148 (111:37) | receive three unsupervised IMT sessions per week, on nonconsecutive days, for 8 weeks | IMT | receive standard-of-care (SOC) treatment | SOC | Mental disorders | Fatigue | 56 | Unclear | Y | N | N | N | / |
| 83 | 2022 | Merchante | RCT, double-blind, phase 2 | Pharmacological intervention | 115 (39:37:39) | receive sarilumab-400 or sarilumab-200 with standard of care | Sarilumab | receive standard-of-care (SOC) treatment | SOC | nmune dysregulatio | Systemic inflammation | 28 | RT-PCR test | N | / | Y | / | / |
| 84 | 2021 | Mohajeri | RCT, open-label | Pharmacological intervention, nutritional supplement | 452 (222:230) | receive 10 g of glutamine supplement three times per day | L-Glutamine | receive standard-of-care (SOC) treatment | SOC | /sregulation and tas | Inflammatory response, oxidative stress, loss of appetite and malnutrition | 5 | thorax CT and RT-PCR test | N | / | N | N | / |
| 85 | 2022 | N. Sharma | RCT, double-blind | Non-pharmacological intervention | 62 (31:31) | receive 50 minutes of yoga intervention along | Yoga | receive standard-of-care (SOC) treatment | SOC | Mental disorders | Stress, anxiety and depression | 18 | RT-PCR test | N | mild/moderate | Y | N | / |
| 86 | 2022 | Nagy | RCT, single-blind | Non-pharmacological intervention, exercise training | 52 (26:26) men | receive manual DR and IMT (POWERbreath) in addition to their prescribed medications | MDRT + IMT | receive IMT training | IMT | Mental disorders | Fatigue | 42 | chest CT scan | Y | N | N | N | / |
| 87 | 2022 | Nambi | RCT, single-blind | Non-pharmacological intervention, exercise training | 76 (38:38) | receive high-intensity aerobic training in addition to standard care for 30 minutes/session, 1 session/day, 4 days/week for 8 weeks | AT_HI | receive low-intensity aerobic training in addition to standard care for 30 minutes/session, 1 session/day, 4 days/week for 8 weeks | AT_LI | Mental disorders | Fatigue | 56 | Unclear | Y | N | N | N | older man |

# Table 1.

| No. | Year | Author | Study design | Intervention type | Sample size | Intervention/Case group | Intervention abbreviation | Control group | Control abbreviation | Symptom cluster1 | Symptom cluster2 | Follow-up (days) | Diagnosis of COVID-19 | Post-COVID-19 (Y/N) | Disease severity (Severe/Moderate/ Mild) | Hospitalized (Y/N) | ICU (Y/N) | Comment |
|---|---|---|---|---|---|---|---|---|---|---|---|---|---|---|---|---|---|---|
| 88 | 2022 | Okan | RCT, open-label | Non-pharmacological intervention, exercise training | 52 (26:26) | receive breathing exercise three times a day for 5 weeks (one session performed via telemedicine each week) | RMT | receive standard-of-care (SOC) treatment | SOC | Mental disorders | Fatigue | 35 | Unclear | Y | N | N | N | / |
| 89 | 2022 | Orendáčová | a longitude observational study | Non-pharmacological intervention, cognitive behavioral therapy (CBT) | 10 | receive NFB training included five NFB sessions which were completed within 2 weeks | NFBT | receive standard-of-care (SOC) treatment | SOC | Mental disorders | Anxiety, fatigue and depression | 14 | antigen or RT-PCR or antibody test | Y | N | N | N | / |
| 90 | 2021 | Özlü | RCT, open-label | Non-pharmacological intervention, exercise training | 67 (33:34) | receive progressive muscle relaxation exercises twice a day for 5 days with the researcher's supervision | PMR | receive standard-of-care (SOC) treatment | SOC | Mental disorders | Anxiety and insomnia | 5 | Unclear | N | / | / | / | / |
| 91 | 2021 | Parizad | RCT, single-blind | Non-pharmacological intervention, cognitive behavioral therapy (CBT) | 110 (55:55) | receive 10 training sessions of guided imagery | GI | receive standard-of-care (SOC) treatment | SOC | Mental disorders | Anxiety | 5 | Unclear | N | N | N | N | / |
| 92 | 2021 | Perepu | RCT, open-label | Pharmacological intervention, anticoagulant therapy | 173 (86:87) | receive standard prophylactic dose enoxaparin | LMWH (Enoxaparin) (AC_PD) | receive intermediate dose enoxaparin | LMWH (Enoxaparin) AC_ID | Cardiovascular symptoms | Thrombosis | 30 | RT-PCR test | N | severe | Y | / | / |
| 93 | 2022 | Philip | RCT, single-blind | Non-pharmacological intervention, exercise training | 150 (74:76) | receive ENO Breathe | ENO-Breathe | receive standard-of-care (SOC) treatment | SOC | Organ damage/failure and mental disorders | Fatigue and anxiety | 42 | Unclear | Y | N | N | N | / |
| 94 | 2022 | Pires | RCT, open-label | Non-pharmacological intervention, olfactory training | 80 (26:54) | receive an advanced olfactory training set (AOT) with 8 essential oils (rose, eucalyptus, clove, lemon, citronella, mint, vanilla, and cedar wood), exposed to each odor for 15-s twice daily, with a 30-s interval between odors | AOT | receive a classical olfactory training set (COT) with 4 essential oils (rose, eucalyptus, clove, and lemon), exposed to each odor for 15-s twice daily, with a 30-s interval between odors | OT | Smell/taste disorder | Olfactory dysfunction | 21 | RT-PCR test | Y | N | N | N | / |
| 95 | 2021 | Pontali | CCT, open-label | Pharmacological intervention, anti-inflammatory treatment (AIT) | 128 (63 (33:30) : 65 (21:44)) | Anakinra + GC group (patients admitted to the emergency room because of the impossibility of starting anakinra treatment immediately): receive starting dose of 100 mg anakinra every 8 hours for 3 days, followed by tapering (100 mg every 12 hours for 1-3 days, followed by 100 mg every 24 hours for 1-3 days) according to the clinical response, for a maximum of 9 days and receive glucocorticoid treatment (intravenous methylprednisolone, 1 to 2 mg/kg once or twice daily, with tapering) as the first drugAnakinra alone group: receive starting dose of 100 mg anakinra every 8 hours for 3 days, followed by tapering (100 mg every 12 hours for 1-3 days, followed by 100 mg every 24 hours for 1-3 days) according to the clinical response, for a maximum of 9 dayssome: receive antivirals (lopinavir/ritonavir [400/100 mg twice daily] or darunavir [800 mg]/ritonavir [100 mg daily] for 7 days, and enoxaparin [4000 IU per day] adapted according to body weight, kidney function, and D-dimer levels) | Anakinra + GC | SOC only group: only receive standard-of-care (SOC) treatment (consisting of hydroxychloroquine (400 mg twice daily on the first day followed by 200-400 mg twice daily for 7 days) and/or azithromycin (500 mg daily for 7 days))SOC + rescue treatment (receive anakinra, intravenous tocilizumab (8 mg/kg to a maximum dose of 800 mg), and glucocorticoids):<br>10 anakinra + GC<br>3 anakinra alone<br>6 tocilizumab alone<br>1 tocilizumab + GC<br>1 GC alonesome: receive antivirals (lopinavir/ritonavir [400/100 mg twice daily] or darunavir [800 mg]/ritonavir [100 mg daily] for 7 days, and enoxaparin [4000 IU per day] adapted according to body weight, kidney function, and D-dimer levels) | SOC | Immune dysregulation | Inflammatory response | 7 | RT-PCR test | N | severe | Y | / | / |
| 96 | 2022 | Ponthieux | a longitude observational study | Pharmacological intervention | 23 (15:8) | receive TCZ | Tocilizumab | receive standard-of-care (SOC) treatment | SOC | Smell/taste disorder | Olfactory dysfunction | 10 | rRT-PCT test | N | severe | Y | / | / |
| 97 | 2021 | Rashid | RCT, double-blind | Pharmacological intervention, intra-nasal spray | 276 (138:138) | receive intranasal betamethasone sodium phosphate drops (0.1 mg/ mL) 3-times daily until recovery for a maximum of one month | IBSPD | receive placebo | Placebo | Smell/taste disorder | Olfactory dysfunction | 30 | RT-PCR test | N | mild/moderate | N | N | / |
| 98 | 2021 | Rivaroxaban | RCT, open-label | Pharmacological intervention, anticoagulant therapy | 318 (159:159) | receive thromboprophylaxis with rivaroxaban 10 mg/day for 35 days | Rivaroxaban (AC_PD) | receive standard-of-care (SOC) treatment | SOC | Cardiovascular symptoms | Thrombosis | 35 | RT-PCR, antigen or IgM test | Y | N | Y | / | / |
| 99 | 2022 | Rohani | RCT, triple-blind | Pharmacological intervention, nutritional supplement | 180 (89:91) | receive 25000 IU/d oral vitamin A for 10 days in addition to the standard treatment recommended by the national protocol | VA | receive the national standard treatment for COVID-19 (hydroxychloroquine) | SOC | Mental disorders | Fatigue | 10 | RT-PCR test | N | / | N | N | / |
| 100 | 2021 | Roozbeh | RCT, double-blind | Pharmacological intervention | 55 (27: 28) | receive sofosbuvir/daclatasvir plus hydroxychloroquine | Sofosbuvir + Daclatasvir | receive hydroxychloroquine | SOC | Mental and taste disorders | Fatigue and loss of appetite | 30 | chest CT scan | N | mild | N | N | / |
| 101 | 2021 | Sadeghi | an observational study | Pharmacological intervention, mesenchymal stem cell (MSC) therapy | 10 | receive DSCs 1–2 times at a dose of 1 × 106/ kg | MSCT | No | No | Organ damage/failure and immune dysregulation | Acute respiratory distress syndrome (ARDS) and inflammatory response | 10 | qRT- PCR | N | severe | Y | / | ARDS patients |
| 102 | 2022 | Sadeghipour | RCT, open-label | Pharmacological intervention, anticoagulant therapy | 375 (187:188) | receive intermediate-dose prophylactic anticoagulation | AC_ID | receive standard-dose prophylactic anticoagulation (SDPAC) | AC_PD | Mental disorders | Functional Impairment and depression | 90 | Unclear | N | severe | / | / | / |
| 103 | 2022 | Saeedi-Boroujeni | RCT, open-label | Pharmacological intervention | 60 (30:30) | receive Tranilast (300 mg daily) in addition to the antiviral drugs for 7 days | Tranilast | receive standard-of-care (SOC) treatment | SOC | Immune dysregulation | Inflammatory response | 21 | RT-PCR test | N | severe | Y | / | / |
| 104 | 2021 | Saleh | CCT, open-label, a pilot trial, phase 1 | Pharmacological intervention, mesenchymal stem cell (MSC) therapy | 5 | receive 3 intravenous injections 3 days apart | MSCT | No | No | nmune dysregulatio | Inflammatory response | 28 | RT-PCR test | N | severe | / | N | / |
| 105 | 2021 | Sayad | RCT, open-label | Pharmacological intervention | 80 (40:40) | receive a fixed-dose combination tablet containing 400 mg sofosbuvir and 100 mg velpatasvir (Shari Pharmaceutical Industry Co., Tehran, Iran) orally once daily for 10 days plus the national standard of care for 10 days | Sofosbuvir + Velpatasvir | receive the national standard of care including 400 mg hydroxychloroquine as a single dose and lopinavir/ritonavir (400 mg/100 mg) orally twice daily | SOC | Organ damage/failure | Respiratory and multi-organ failure | 28 | RT-PCR test | N | moderate/severe | Y | / | / |

Table 1.

| No. | Year | Author | Study design | Intervention type | Sample size | Intervention/Case group | Intervention abbreviation | Control group | Control abbreviation | Symptom cluster1 | Symptom cluster2 | Follow-up (days) | Diagnosis of COVID-19 | Post-COVID-19 (Y/N) | Disease severity (Severe/Moderate/ Mild) | Hospitalized (Y/N) | ICU (Y/N) | Comment |
|---|---|---|---|---|---|---|---|---|---|---|---|---|---|---|---|---|---|---|
| 106 | 2020 | Sekhavati | RCT, open-label | Pharmacological intervention | 111 (56:55) | receive oral AZM 500 mg daily, oral LPV/r 400/100 mg twice daily and oral HCQ 400 mg daily for 5 days | Sofosbuvir + Velpatasvir | receive oral LPV/r 400/100 mg twice daily and oral HCQ 400 mg daily; for both treatment groups | SOC | Overall | Overall | 30 | RT-PCR test and chest CT scan | N | / | Y | / | / |
| 107 | 2021 | Shabahang | RCT, open-label | Non-pharmacological intervention, cognitive behavioral therapy (CBT) | 150 (75:75) | receive a video-based cognitive behavioral therapy (CBT) consisting of nine 15-20-minute sessions 3 days a week for 3 weeks | vCBT | receive standard-of-care (SOC) treatment | SOC | Mental disorders | Anxiety | 21 | Unclear | N | / | / | N | / |
| 108 | 2022 | Shah | an observational study | Pharmacological intervention, immune-modulator | 315 (105:210) | receive MIP at a dosage of 0.1 ml in-tramuscular three times a day at three different sites for 3 consecutive days in addition to BST | Mycobacterium indicus pranii | receive standard-of-care (SOC) treatment | SOC | Immune dysregulation | Inflammatory response | 5 | RT-PCR test and chest CT scan | N | severe | Y | / | / |
| 109 | 2022 | Shome | CCT, open-label, a before-and-after self-controled trial | Pharmacological intervention | 20 females | receive a dose of 1.5 mL intra-dermal administration of QR678 Neo(®) hair growth factor formulation in the scalp in each session; a total of 8 sessions; each session 4 weeks apart | Azithromycin | No | No | Skin symptoms | Persistent Telogen Effluvium (TE) (Hair loss) | 224 | RT-PCR test | Y | N | N | N | / |
| 110 | 2020 | Shu | RCT, open-label | Pharmacological intervention, mesenchymal stem cell (MSC) therapy | 41 (12:29) | receive a standard treatment plus umbilical cord mesenchymal stem cell infusion | MSCT | receive standard-of-care (SOC) treatment | SOC | Immune dysregulation | Inflammatory response | 28 | RT-PCR test and chest CT scan | N | severe | / | / | / |
| 111 | 2021 | Silveira | RCT, open-label | Pharmacological intervention, adjunct treatment | 124 (42:40:42) | receive Propomax® capsules produced with dehydrated Standardized Brazilian Green Propolis Extract, EPP-AF® for 7 days at 800 mg/day (two 100 mg capsules, four times a day) plus standard care, or 400 mg/day (one 100 mg capsule, four times a day) plus standard care | Propolis | receive standard-of-care (SOC) treatment | SOC | Immune dysregulation | Inflammatory response | 28 | Unclear | N | moderate/severe | Y | / | / |
| 112 | 2022 | Song | RCT, double-blind, phase 3 | Pharmacological intervention | 22 (10:12) | receive a single dose of CD24Fc antibody (480 mg IV infusion) | CD24Fc | receive placebo | Placebo | Immune dysregulation | Inflammatory response | 28 | RT-PCR test | N | severe | Y | / | / |
| 113 | 2021 | Spyropoulos | RCT, double-blind | Pharmacological intervention, anticoagulant therapy | 253 (129:124) | receive enoxaparin at a dose of 1 mg/kg subcutaneously twice daily if CrCl was 30 mL/min/1.73 m2 or greater or 0.5 mg/kg twice daily if CrCl was 15-29 mL/min/1.73 m2 | LMWH (Enoxaparin) (AC_TD) | receive prophylactic or intermediate-dose heparin regimens per local institutional standard | Heparin (AC_PD/AC_ID) | rdiovascular symptc | Thrombosis | 100 | Unclear | N | severe | Y | / | / |
| 114 | 2022 | Stockmann | RCT, open-label | Pharmacological intervention | 49 (23:26) | receive CytoSorb for 3–7 days | CytoSorb | receive standard-of-care (SOC) treatment | SOC | symptoms and orgai | Vasoplegic shock and multi-organ failure | 30 | RT-PCR test | N | severe | Y | Y | / |
| 115 | 2021 | Thomas | RCT, open-label | Pharmacological intervention, nutritional supplement | 214 (58:58:48:50) | receive 10 days of zinc gluconate (50 mg) and ascorbic acid (8000 mg) | Zinc + VC | receive 10 days of zinc gluconate (50 mg), ascorbic acid (8000 mg), or standard-of-care (SOC) treatment | SOC | Mental disorders | Fatigue | 28 | RT-PCR test | N | / | N | N | / |
| 116 | 2020 | Tomazini | RCT, open-label | Pharmacological intervention | 299 (151:148) | receive 20 mg of dexamethasone intravenously daily for 5 days, 10 mg of dexamethasone daily for 5 days or until ICU discharge, plus standard care | Dexamethasone | receive standard-of-care (SOC) treatment | SOC | Organ damage/failure | Acute respiratory distress syndrome (ARDS) | 28 | RT-PCR test | N | moderate/severe | Y | Y | / |
| 117 | 2021 | Vaira | a case-control study | Pharmacological intervention | 18 (9:9) | receive corticosteroid therapy | Dexamethasone | receive standard-of-care (SOC) treatment | SOC | Smell/taste disorder | Olfactory dysfunction | 15 | RT-PCR test | Y | mild/moderate | N | N | / |
| 118 | 2020 | Valizadeh | RCT, double-blind | Pharmacological intervention | 40 (20:20) | receive 160 mg of Nano-curcumin in four 40 mg capsules daily for 14 days | Nano-curcumin | receive placebo | Placebo | nmune dysregulatio | Inflammatory response | 14 | RT-PCR test | N | / | Y | Y | / |
| 119 | 2022 | Vigstedt | RCT, single-blind | Pharmacological intervention, intravenous injection | 80 (41:39) | receive prostacyclin infusion (1 ng/kg/min) | Epoprostenol | receive placebo | Placebo | nmune dysregulatio | Inflammatory response | 1 | Unclear | N | severe | Y | / | mechanically ventilated patients |
| 120 | 2020 | Vink | RCT, open-label | Non-pharmacological intervention, cognitive behavioral therapy (CBT) | 156 (52:52:52) | receive cognitive behavioural therapy (CBT) | CBT | receive doxycycline or placebo | Placebo | Mental disorders | Fatigue and functional impairment | 364 | RT-PCR test | Y | N | N | N | / |
| 121 | 2021 | Wang | RCT, open-label | Pharmacological intervention, monoclonal antibody | 65 (34:31) | receive tocilizumab in addition to standard care | Tocilizumab | receive standard-of-care (SOC) treatment | SOC | Organ damage/failur | Hypoxaemia | 14 | RT-PCR test | N | moderate/severe | / | N | / |
| 122 | 2020 | Wu | an observational study, phase 1 | Pharmacological intervention, embryonic stem cell therapy | 27 | receive intravenous transfusion of hESC-IMRCs at a dose of 3 × 106 cells/kg of body weight | MSCT | No | No | Organ damage/failur | Pulmonary fibrosis | 84 | chest CT scan | N | moderate/severe | / | N | / |
| 123 | 2021 | Xia | RCT, double-blind | Pharmacological intervention, traditional Chinese medicine (TCM) treatment | 76 (43:33) | receive AVD plus SFJDC | SFJDC | receive AVD | SOC | Mental disorders | Fatigue | 21 | rRT-PCT test | N | moderate | Y | N | / |
| 124 | 2022 | Y. Li | RCT, open-label | Pharmacological intervention, traditional Chinese medicine (TCM) treatment | 346 (106:106:106) | receive treatment with QFPD-LHQW | QFPDC + LHQWC | receive treatment with LHQW or QFPD | QFPDC/LHQWC | nmune dysregulatio | Inflammatory response | 30 | rRT-PCR test | N | / | Y | N | / |
| 125 | 2021 | Z. Liu | RCT, open-label | Non-pharmacological intervention, cognitive behavioral therapy (CBT) | 101 (51:50) | receive cCBT + TAU treatment for 1 week | cCBT | receive TAU treatment | SOC | Mental disorders | Anxiety, depression and insomnia | 7 | Unclear | N | / | Y | N | / |
| 126 | 2022 | Zasadzka | RCT, open-label | Non-pharmacological intervention, exercise training | 28 (14:14) | receive additional training using an EMG rehabilitation robot | EMG-RRT | receive standard-of-care (SOC) treatment | SOC | Mental disorders | Fatigue | 42 | Unclear | Y | severe | Y | Y | / |
| 127 | 2022 | Zheng | RCT, open-label | Non-pharmacological intervention, cognitive behavioral therapy (CBT) | 70 (35:35) | receive positive therapy and full perfusion therapy based on conventional therapy | Aromatherapy | receive standard-of-care (SOC) treatment | SOC | Mental disorders | Posttraumatic stress disorder (PTSD), anxiety and depression | 30 | Unclear | Y | N | Y | N | / |
| 128 | 2020 | Zhou | RCT, open-label | Pharmacological intervention | 77 (46:7:24) | receive antiviral treatment with a combination of IFN-α2b plus ARB | IFN-α2b + ARB | receive antiviral treatment with either IFN-α2b or ARB (arbidol hydrochloride) | IFN-α2b/ARB | nmune dysregulatio | Inflammatory response | 42 | RT-PCR test | N | severe | Y | N | / |
| 129 | 2022 | Zilberman-Itskovich | RCT, open-label | Non-pharmacological intervention, hyperbaric oxygen therapy (HBOT) | 73 (37:36) | receive hyperbaric oxygen therapy (HBOT) | HBOT | receive standard-of-care (SOC) treatment | SOC | Immune dysregulation | Neuroinflammatio n and hypercoagulability | 21 | RT-PCR test | Y | N | / | N | / |